\pdfoutput=1

\documentclass[11pt,twoside,a4paper,cmspaper,final,collab]{cms-tdr}
\def\svnVersion{267485}\def\cmsCernNo{2013-037}\def\cmsCernDate{\today}\def\cmsMessage{Published in Physics Letters B as \href{http://dx.doi.org/10.1016/j.physletb.2012.02.085}{\doi{10.1016/j.physletb.2012.02.085.}}}

\begin{document}\cmsNoteHeader{HIG-11-031}

\hyphenation{had-ron-i-za-tion}
\hyphenation{cal-or-i-me-ter}
\hyphenation{de-vices}

\newcommand\T{\rule{0pt}{2.3ex}}
\newcommand\B{\rule[-1.0ex]{0pt}{0pt}}
\newcommand\Mbb   {\ensuremath{m(\mathrm{jj})}}
\newcommand\Mjj     {\ensuremath{m(\mathrm{jj})}}

\newcommand\BDT {\textsc{bdt}}
\newcommand\Zj  {\ensuremath{\cPZ+\text{jets}}}
\newcommand\Wj  {\ensuremath{\PW+\text{jets}}}
\newcommand\Vj  {\ensuremath{V+\text{jets}}}
\newcommand\VtoBB {\ensuremath{V\to\bbbar}}
\newcommand\mH{\ensuremath{m_\PH}}

\renewcommand\HT {\ensuremath{\PH_\mathrm{T}}}
\newcommand\WZ  {\ensuremath{\PW\cPZ}}
\newcommand\WW  {\ensuremath{\PW\PW}}
\newcommand\ZZ  {\ensuremath{\cPZ\cPZ}}
\newcommand\Wlf   {\ensuremath{{\PW}lf}}
\newcommand\Zlf   {\ensuremath{{\cPZ}lf}}

\newcommand\V {\ensuremath{V}}
\newcommand\ZnnH  {\ensuremath{\cPZ(\cPgn\cPgn)\PH}}
\newcommand\ZllH  {\ensuremath{\cPZ(\ell\ell)\PH}}
\newcommand\ZmmH  {\ensuremath{\cPZ(\Pgm\Pgm)\PH}}
\newcommand\ZeeH  {\ensuremath{\cPZ(\Pe\Pe)\PH}}
\newcommand\ZH   {\ensuremath{\cPZ\PH}}
\newcommand\WH   {\ensuremath{\PW\PH}}
\newcommand\VH   {\ensuremath{V\PH}}
\newcommand\WlnH  {\ensuremath{\PW(\ell\cPgn)\PH}}
\newcommand\WmnH  {\ensuremath{\PW(\Pgm\cPgn)\PH}}
\newcommand\WenH  {\ensuremath{\PW(\Pe\cPgn)\PH}}
\newcommand\WtoLN {\ensuremath{\PW\to\ell\cPgn}}
\newcommand\WtoEN {\ensuremath{\PW\to\Pe\cPgn}}
\newcommand\WtoMN {\ensuremath{\PW\to\Pgm\cPgn}}
\newcommand\ZtoBB {\ensuremath{\cPZ\to\bbbar}}
\newcommand\ZtoNN {\ensuremath{\cPZ\to\cPgn\cPagn}}
\newcommand\ZtoLL {\ensuremath{\cPZ\to\ell\ell}}
\newcommand\ZtoMM {\ensuremath{\cPZ\to\MM}}
\newcommand\ZtoEE {\ensuremath{\cPZ\to\EE}}
\newcommand\ZmmJ  {\ensuremath{\cPZ(\Pgm\Pgm)+\text{jets}}}
\newcommand\ZnnJ  {\ensuremath{\cPZ(\cPgn\bar{\cPgn})+\text{jets}}}
\newcommand\WJ  {\ensuremath{\PW + \text{jets}}}
\newcommand\HBB   {\ensuremath{\PH\to\bbbar}}
\newcommand\HTT   {\ensuremath{\PH\to\TT}}
\newcommand\mtW   {\ensuremath{M_{\mathrm{T}}}}
\newcommand\ptl   {\ensuremath{p_{\mathrm{T}}^{\ell}}}
\newcommand\MyZ   {\ensuremath{\cPZ}}
\newcommand\MyW   {\ensuremath{\PW}}
\newcommand\MyH   {\ensuremath{\PH}}
\newcommand\Vudscg   {\ensuremath{V+\cPqu\cPqd\cPqs\cPqc\Pg}}
\newcommand\Wudscg   {\ensuremath{\PW+\cPqu\cPqd\cPqs\cPqc\Pg}}
\newcommand\Wenudscg   {\ensuremath{\PW(\Pe\cPgn)+\cPqu\cPqd\cPqs\cPqc\Pg}}
\newcommand\Wmnudscg   {\ensuremath{\PW(\Pgm\cPgn)+\cPqu\cPqd\cPqs\cPqc\Pg}}
\newcommand\Wenbb   {\ensuremath{\PW(\Pe\cPgn)+\bbbar}}
\newcommand\Wmnbb   {\ensuremath{\PW(\Pgm\cPgn)+\bbbar}}
\newcommand\Zeebb   {\ensuremath{\cPZ(\Pe\Pe)+\bbbar}}
\newcommand\Zmmbb   {\ensuremath{\cPZ(\Pgm\Pgm)+\bbbar}}
\newcommand\Zudscg   {\ensuremath{\cPZ+\cPqu\cPqd\cPqs\cPqc\Pg}}
\newcommand\Zeeudscg   {\ensuremath{\cPZ(\Pe\Pe)+\cPqu\cPqd\cPqs\cPqc\Pg}}
\newcommand\Zmmudscg   {\ensuremath{\cPZ(\Pgm\Pgm)+\cPqu\cPqd\cPqs\cPqc\Pg}}
\newcommand\Zenbb   {\ensuremath{\cPZ(\Pe\cPgn)+\bbbar}}
\newcommand\Zmnbb   {\ensuremath{\cPZ(\Pgm\cPgn)+\bbbar}}
\newcommand\Wbb   {\ensuremath{\PW\bbbar}}
\newcommand\Zbb   {\ensuremath{\cPZ\bbbar}}
\newcommand\Vbb   {\ensuremath{V\bbbar}}
\newcommand\Zll   {\ensuremath{\cPZ(\ell\ell)}}
\newcommand\ptjj   {\ensuremath{{\pt}(\mathrm{jj})}}
\newcommand\MZ {\ensuremath{M_{\cPZ}}}
\newcommand\dRJJ {\ensuremath{\Delta R(\mathrm{j1,j2})}}
\newcommand\dEtaJJ {\ensuremath{|\Delta \eta(\mathrm{jj})|}}
\newcommand\dphiVH {\ensuremath{\Delta\phi(V,\PH)}}
\newcommand\dphiWH {\ensuremath{\Delta\phi(\PW,\PH)}}
\newcommand\dphiZH {\ensuremath{\Delta\phi(\cPZ,\PH)}}
\newcommand\dphiMJ {\ensuremath{\Delta\phi(\MET,\text{jet})}}
\newcommand\cosTH {\ensuremath{\cos{\theta^*}}}
\newcommand\dThPull {\ensuremath{\Delta\theta_{\text{pull}}}}
\newcommand\ptV {\ensuremath{\PT(V)}}
\newcommand\ptH {\ensuremath{\PT(\PH)}}
\newcommand\ptZ {\ensuremath{\PT(\cPZ)}}
\newcommand\ptW {\ensuremath{\PT(\PW)}}
\newcommand\Naj {\ensuremath{N_{\mathrm{aj}}}}
\newcommand\Nal {\ensuremath{N_{\mathrm{al}}}}
\newcommand\etaTF {\ensuremath{\left | \eta \right | < 2.5}}
\newcommand\Bexp {\ensuremath{B_{\text{exp}}}}
\newcommand\Bobs {\ensuremath{B_{\text{obs}}}}
\newcommand\Nobs {\ensuremath{N_{\text{obs}}}}

\RCS$Revision: 107575 $
\RCS$HeadURL: svn+ssh://svn.cern.ch/reps/tdr2/papers/HIG-11-031/trunk/HIG-11-031.tex $
\RCS$Id: HIG-11-031.tex 107575 2012-02-28 13:28:42Z jacobo $
\newlength\cmsFigWidth
\ifthenelse{\boolean{cms@external}}{\setlength\cmsFigWidth{0.98\columnwidth}}{\setlength\cmsFigWidth{0.75\textwidth}}
\ifthenelse{\boolean{cms@external}}{\providecommand{\cmsLeft}{top}}{\providecommand{\cmsLeft}{left}}
\ifthenelse{\boolean{cms@external}}{\providecommand{\cmsRight}{bottom}}{\providecommand{\cmsRight}{right}}
\cmsNoteHeader{HIG-11-031} 

\title{Search for the standard model Higgs boson decaying to bottom quarks in \Pp\Pp\ collisions at $\sqrt{s}=7$\TeV}

\date{\today}

\abstract{
   A search for the standard model Higgs boson (\PH) decaying to
\bbbar when produced in association with weak
   vector bosons
   (\V) is reported for the following modes: $\PW(\Pgm\cPgn)\PH,$ $\PW(\Pe\cPgn)\PH,$
   $\cPZ(\Pgm\Pgm)\PH,$ $\cPZ(\Pe\Pe)\PH$ and
   $\cPZ(\nu\nu)\PH$. The search is performed in a data
   sample corresponding to an integrated luminosity of 4.7\fbinv, recorded by the CMS detector
   in proton-proton collisions at the LHC with a center-of-mass energy
   of 7\TeV.
   No significant excess of events above the
   expectation from background is observed. Upper limits on the
  $\V\PH$ production cross section times the $\PH\to \bbbar$
  branching ratio, with respect to the expectations for a standard model Higgs boson, are derived for a Higgs boson in the mass range
   110--135\GeV. In this range, the observed 95\% confidence level upper limits
   vary from 3.4 to 7.5 times the standard model prediction; the
   corresponding expected limits vary from 2.7 to 6.7 times the
   standard model prediction.
}

\hypersetup{%
pdfauthor={CMS Collaboration},%
pdftitle={Search for the standard model Higgs boson decaying to bottom quarks in pp collisions at sqrt(s)=7 TeV},%
pdfsubject={CMS},%
pdfkeywords={CMS, physics, Higgs}}

\maketitle 

\section{Introduction}\label{sec:intro}
\label{sec:introduction}

The process by which the electroweak symmetry is broken in nature remains elusive.
In the standard model (SM)~\cite{SM1,SM2,SM3} the Higgs mechanism is
considered to be the
explanation~\cite{Englert:1964et,Higgs:1964ia,Higgs:1964pj,Guralnik:1964eu,Higgs:1966ev,Kibble:1967sv}.
The search for the Higgs boson is currently one of the most important endeavors of experimental particle physics.

Direct searches by experiments at the Large Electron-Positron Collider (LEP)
have set a $95\%$ confidence level (CL) lower bound on the Higgs boson mass
of $\mH > 114.4$\GeV~\cite{LEPHIGGS}. Direct searches at the Tevatron
exclude at $95\%$ CL the $162$--$166$\GeV mass
range~\cite{TEVHIGGS_2010}, and the ATLAS experiment at the Large
Hadron Collider (LHC) excludes, also at 95\% CL, the following three regions:
$\mH \notin$~145--206, 214--224, and 340--450\GeV~\cite{ATLAS:2011aa,
Aad:2011uq, ATLAS:2011af}. Measurements of the \PW\ boson and top quark masses at LEP
and the Tevatron, combined with precision measurements of
electroweak parameters at the \cPZ\ pole, provide an indirect constraint
of $\mH < 158$\GeV at $95\%$ CL~\cite{LEPewkfits}. The most likely mass for the SM
Higgs boson remains near the LEP limit, where the
Higgs boson decays
predominantly into \bbbar. Experiments at the Tevatron have set 95\%
CL upper limits on the production cross section for a Higgs boson in this low-mass region.
These limits range from approximately 4 to 10 times the standard model prediction,
depending on the channels
studied~\cite{Aaltonen:2009dh,Aaltonen:2009jg,Aaltonen:2010pa,Aaltonen:2011dh,Abazov20116,AbazovZnnHbb,AbazovZllHbb}. The
observation of the $\PH\to \bbbar$ decay is
of great importance in determining the nature of the Higgs boson.

At the LHC the main SM Higgs boson production mechanism is gluon fusion,
with a cross section of ${\approx}17$\unit{pb} for $\mH=120$\GeV~\cite{Ellis:1975ap,Georgi:1977gs,Djouadi:1991tka,Dawson:1990zj,Spira:1995rr,Harlander:2002wh,Anastasiou:2002yz,Ravindran:2003um,Catani:2003zt,Aglietti:2004nj,Degrassi:2004mx,Baglio:2010ae,Actis:2008ug,Anastasiou:2008tj,deFlorian:2009hc,Djouadi:1997yw,LHCXSWG}.
However, in this production mode, the detection of the $\PH\to \bbbar$ decay
is considered nearly impossible due to overwhelming dijet production expected from quantum-chromodynamic (QCD) interactions.
The same holds true for the next most copious production mode, through
vector-boson fusion, with a cross section of ${\approx} 1.3$\unit{pb}~\cite{Ciccolini:2007jr,Ciccolini:2007ec,Figy:2003nv,Arnold:2008rz,Bolzoni:2010xr}. Processes
in which a low-mass Higgs boson is produced in association with a vector boson~\cite{PhysRevD.18.1724} have
cross sections of ${\approx}0.66$\unit{pb} and ${\approx}0.36$\unit{pb} for \PW\PH\ and \cPZ\PH,
respectively.

In this Letter a search for the standard model Higgs
boson in the $\Pp\Pp\to\V\PH$ production mode is presented, where \V\
is either a \PW\ or a \cPZ\ boson. The analysis
is performed in the  $110$--$135$\GeV Higgs boson mass range, using a data sample corresponding to an integrated luminosity of
 4.7\fbinv, collected in 2011 by the Compact Muon Solenoid (CMS) experiment at a
center-of-mass energy of 7\TeV. The following final states are included:
$\PW(\Pgm\cPgn)\PH$, $\PW(\Pe\cPgn)\PH$, $\cPZ(\Pgm\Pgm)\PH$,
$\cPZ(\Pe\Pe)\PH$ and $\cPZ(\cPgn\cPgn)\PH$, all with the Higgs boson decaying to \bbbar. Backgrounds arise from production of \PW\ and \cPZ\ bosons in association with
jets (from all quark flavors), singly and pair-produced top quarks (\ttbar), dibosons and
QCD multijet processes.
Simulated samples of signal and backgrounds are used to provide
guidance in the optimization of the analysis as a function of the
Higgs boson
mass.  Control regions in data are selected to adjust the simulations and
estimate the contribution of the main backgrounds in the signal
region. Upper limits at the 95\% CL on the  $\Pp\Pp\to\VH$ production cross section are
obtained for Higgs boson masses between
110--135\GeV. These limits are based on the observed event count and
background estimate in signal-enriched regions selected using the output
discriminant of a boosted-decision-tree algorithm~\cite{Roe:2005hm} (\BDT\
analysis). As a cross-check, limits are also derived from the
observed event count in the invariant mass distribution of $\PH\to
\bbbar$ candidates (\Mbb\ analysis).

\section{CMS Detector and Simulations}\label{sec:detector}
A detailed description of the CMS detector can be found
elsewhere~\cite{CMSDETECTOR}.
The momenta of charged particles are measured using a silicon pixel
and strip tracker that covers the pseudorapidity range
$|\eta|\le 2.5$ and is immersed in a 3.8\unit{T}
solenoidal magnetic field. The pseudorapidity is defined as $\eta = -\ln(\tan(\theta/2))$, where $\theta$
is the polar angle of the trajectory of a particle with respect to
the direction of the counterclockwise proton beam.
Surrounding the tracker are a crystal electromagnetic calorimeter
(ECAL) and a brass-scintillator hadron calorimeter (HCAL), both used to
measure particle energy depositions and consisting of a barrel assembly and two endcaps. The ECAL
and HCAL extend to a pseudorapidity range of $|\eta|\le 3.0$. A steel/quartz-fiber Cherenkov forward detector (HF) extends the calorimetric
coverage to $|\eta|\le 5.0$. The outermost component of the CMS detector is the
muon system consisting of gas detectors placed in the steel return yoke
to measure the momentum of muons traversing the detector.

Simulated samples of signal and backgrounds are produced using
various event generators, with the CMS detector response modeled
with \GEANTfour~\cite{GEANT4}. The Higgs boson signal samples are
produced using \POWHEG~\cite{POWHEG} interfaced with the \HERWIG~\cite{HERWIG}
event generator. The diboson samples are generated with {\sc pythia 6.4}~\cite{Pythia}.
The \MADGRAPH 4.4~\cite{MadGraph} generator is used for the \PW+jets, \cPZ+jets, and \ttbar samples.
The single-top samples are produced with \POWHEG and the QCD
multijet samples with \PYTHIA. The default set of parton distribution
functions (PDF) used to produce these samples is
CTEQ6L1~\cite{CTEQ6L1}. The
\PYTHIA parameters for the underlying event are set to the Z2 tune~\cite{1107.0330}.

 During the period in which the data for this analysis was recorded, the
LHC instantaneous luminosity reached up to $3.5\times 10^{33}\percms$ and
the average number of \Pp\Pp\ interactions per bunch crossing was approximately
ten. Additional \Pp\Pp\ interactions overlapping with the event of interest in the same bunch crossing,
denoted as  pile-up events (PU), are therefore added in the simulated samples to represent the PU
distribution measured in data.

\section{Triggers and Event Reconstruction}\label{sec:trigger_event}

\subsection{Triggers}

Several triggers are used to collect events consistent with
the signal hypothesis in each of the five channels. For the \PW\PH\ channels the trigger paths consist of several single-lepton
triggers with tight lepton identification. Leptons are also required
to be isolated from other tracks and calorimeter energy depositions to maintain an acceptable trigger
rate. For the \WmnH\ channel, the trigger thresholds for the muon transverse momentum, \PT, are in the range of
 17 to 40\GeV. The higher
thresholds are used for the periods of higher instantaneous luminosity.
The combined trigger efficiency is ${\approx}90\%$ for
signal events that would pass all offline requirements, described
in Section~\ref{sec:event-sel}.
For the \WenH\ channel, the electron \PT\ threshold ranges from 17 to 30\GeV.
The lower-threshold trigger paths require two jets and a minimum
requirement on an online estimate of the missing transverse energy, evaluated in the high level trigger algorithm
as the modulus of the negative vector sum of the transverse momenta
of all reconstructed jets identified by a
particle-flow algorithm~\cite{PFT-09-001}. These extra requirements
help to maintain acceptable trigger
rates during the periods of high instantaneous luminosity.
The combined efficiency for these triggers for signal events that
pass the final offline selection criteria is $>$95\%.

The \ZmmH\ channel uses the same single-muon triggers as the \WmnH\ channel.
For the \ZeeH\ channel, dielectron triggers with lower \PT\
thresholds ($17$ and $8$\GeV ) and tight isolation requirements are used.
These triggers are ${\approx}99\%$ efficient for all \ZH\  signal events that pass the
final offline selection criteria. For the \ZnnH\ channel, a combination of four triggers is used.
The first one requires missing transverse energy $>150$\GeV and is used for the complete
dataset. The other triggers use lower thresholds on the missing
transverse energy (evaluated for
these cases using all energy
deposits in the calorimeter), but require the presence of jets.
One of these triggers requires missing transverse energy above $80$\GeV and a central
($|\eta|<2.4$) jet with \PT\ above 80\GeV, and the other two require the
presence of two central jets with \PT $>20$\GeV and missing
transverse energy thresholds
of 80 and 100\GeV, depending on the luminosity. The combined
trigger efficiency for \ZnnH\ signal events is ${\approx}98\%$ with
respect to the offline event reconstruction and selection, described below.

\subsection{Event reconstruction}
The reconstructed interaction vertex with the largest value of
$\sum_i {\pt}_i^2$, where ${\pt}_i$ is the transverse momentum of
the $i$-th track associated to the vertex, is selected as the primary event vertex. This vertex is used as the
reference vertex
for all relevant objects in the event, which are reconstructed with
the particle-flow algorithm. The PU interactions affect jet momentum
reconstruction, missing transverse energy reconstruction,
lepton isolation and b-tagging efficiency. To mitigate these effects,
a track-based algorithm that filters all  charged hadrons that do not
originate from the primary interaction is used. In addition, a
calorimeter-based algorithm evaluates the energy density in the
calorimeter from interactions not related to the primary vertex and
subtracts its contribution to reconstructed jets in the event~\cite{Cacciari:subtraction}.

Jets are reconstructed from particle-flow objects~\cite{PFT-09-001} using the
anti-$k_T$ clustering algorithm~\cite{antikt},  as implemented in the \textsc{fastjet}
package~\cite{Cacciari:fastjet1,Cacciari:fastjet2},
using a distance parameter of $0.5$.  Each jet is required to
be within $\left | \eta \right | < 2.5$, to have at least two tracks associated to it,
and to have electromagnetic and hadronic energy fractions of at least $1\%$ of the total jet
energy.  Jet energy corrections, as a function of pseudorapidity and
transverse energy of the jet, are applied~~\cite{cmsJEC}. The
missing transverse energy vector is calculated offline as the negative
of the vectorial sum of transverse momenta of all particle-flow objects identified in the
event, and the magnitude of this vector is referred to as \MET in the
rest of this Letter.

Electron reconstruction requires the matching
of an energy cluster in the ECAL with a track in the silicon
tracker~\cite{CMS-PAS-EGM-10-004}. Identification criteria based on the ECAL
shower shape, track-ECAL cluster matching, and consistency with the
primary vertex are imposed.  Additional requirements are imposed to remove electrons
produced by photon conversions. In this analysis, electrons are
considered in the pseudorapidity range $\left | \eta \right | < 2.5$,
excluding the  $1.44 < |\eta| < 1.57$ transition
region between the ECAL barrel and endcap.

Muons are reconstructed using two algorithms~\cite{CMS-PAS-MUO-10-002}: one in which
tracks in the silicon tracker are matched to signals in the muon
chambers, and another in which a global track fit is performed seeded by
signals in the muon system.  The muon
candidates used in the analysis are required to be reconstructed
successfully by both algorithms.  Further identification criteria  are
imposed on the muon candidates to reduce the fraction
of tracks misidentified as muons. These include the number of measurements in the tracker and
the muon system,
the fit quality of the muon track, and its consistency with the primary
vertex.

Charged leptons from \PW\ and \cPZ\ boson decays are expected to be isolated
from other activity in the event. For each lepton candidate, a cone
is constructed around the track direction at the event vertex.  The scalar
sum of the transverse energy of each reconstructed
particle compatible with the primary vertex and contained within the cone is calculated
excluding the contribution from the lepton candidate itself. If this
sum exceeds approximately 10\% of the candidate $\pt$ the lepton is
rejected; the exact requirement depends on the lepton $\eta$, $\pt$ and
flavor.

The Combined Secondary Vertex (CSV) b-tagging
algorithm~\cite{CMS-PAS-BTV-09-001} is used to identify jets that are likely to
arise from the hadronization of b quarks.  This algorithm combines
the information about track impact parameters and secondary
vertices within jets in a likelihood discriminant to provide separation
of b jets from jets originating from light quarks and gluons, and also
from charm quarks. Several working points for the CSV output discriminant are used in the analysis, with
different efficiencies and misidentification rates for b jets.  For a CSV $>0.90$ requirement the efficiencies to tag
b quarks, c quarks, and light quarks, are approximately 50\%,
6\%, and 0.15\%, respectively~\cite{CMS-PAS-BTV-11-003}. The corresponding efficiencies for
CSV $>0.244$ are approximately 82\%, 40\%, and 12\%.

 All events from data and from the simulated samples are required to
pass the same trigger and event reconstruction algorithms. Scale
factors that account for the differences in the performance of these
algorithms between data and simulations are computed and used in
the analysis.

\section{Event Selection}\label{sec:event-sel}

The background processes to \VH\ production are vector-boson+jets, \ttbar,
single-top, dibosons (\V\V)
and QCD multijet production. These overwhelm the signal by several orders of
magnitude. The event selection for the \BDT\ analysis is based first on the kinematic
reconstruction of the vector bosons and the Higgs boson decay
into two b-tagged jets. Backgrounds are then substantially reduced by
requiring a significant boost in the \pt of the vector boson and the
Higgs boson~\cite{PhysRevLett.100.242001}, which can recoil away from each other with a large azimuthal opening angle,  \dphiVH,
between them.
The boost requirements in the \ZllH\ and  WH analyses are
        $\pt>100$ and $\pt>150$\GeV, respectively.  The
        fractions of signal events that satisfy these requirements are approximately
        25\% and 10\%.  For the \ZnnH\ analysis the boost requirement is
        $\pt>160$\GeV.

Candidate \WtoLN\ decays are identified by requiring
the presence of a single isolated lepton and additional missing
transverse energy. Muons are required to have a \pt\ above
20\GeV; the corresponding value for electrons is 30\GeV. For the \WenH\ analysis, \MET is required to be greater than 35\GeV
to reduce contamination from QCD multijet processes. 

Candidate \ZtoLL\ decays are reconstructed by combining
isolated, oppositely charged pairs of electrons or muons, each lepton with \pt$>20$\GeV, and requiring the dilepton invariant
mass to satisfy $75\GeV<m_{\ell\ell}<105\GeV$. The identification of \ZtoNN\ decays
requires \MET$>160$\GeV. The high threshold is dictated by the
trigger and is consistent with a significant boost in the \pt of the \cPZ\
boson. The QCD multijet
background is greatly reduced in this channel when requiring that the \MET does not originate from
mismeasured jets. To that end, a $\dphiMJ>0.5$ radians requirement is applied on the azimuthal angle
between the \MET direction and
the closest jet with \pt$>20$\GeV and $|\eta|<2.5$.
To reduce backgrounds from \ttbar and \WZ\ in the \WH\ and \ZnnH\ channels,
events with additional isolated leptons, \Nal, with \pt$>20$\GeV are rejected.

The reconstruction of the \HBB\ decay is made by
requiring the presence of two central ($|\eta|<2.5$) jets above a minimum \pt threshold, and tagged by the
CSV algorithm. If more than two such jets are
found in the event, the pair of jets with the
highest total dijet transverse momentum, \ptjj, is selected. After the
b-tagging requirements are applied,
the fraction of \HBB\ candidates in signal events that contain the two b jets from the Higgs
boson decay is near 100\%. The background from $\V+\text{jets}$ and
dibosons is reduced significantly
through b tagging, and sub-processes where the two jets originate from
genuine b quarks dominate the final selected data sample.

The \BDT\ analysis is implemented in the TMVA framework~\cite{tmva}. To better separate signal from background under
different Higgs boson mass hypotheses, the \BDT\ is trained separately at
each mass value using simulated samples for signal and background
that pass the event selection described above. The final set of input
variables is chosen by iterative optimization from a larger number of
potentially discriminating variables. The same set is used for all
modes and for all Higgs boson mass hypotheses tested. These include the dijet invariant
mass \Mjj, the dijet transverse momentum \ptjj, the separation in
pseudorapidity between the two jets \dEtaJJ,
the transverse momentum of the vector boson \ptV, the maximum and
minimum CSV values among the two
jets, the azimuthal
angle between the vector boson and the dijets \dphiVH, and the number of additional
central jets \Naj. A signal region, where observed and expected events
are counted, is identified in the \BDT\ output distribution by
optimizing a figure of merit that takes into account
the level of systematic uncertainty on the expected background.

Table~\ref{tab:CC_BDT_sel} summarizes the selection criteria used in
each of the five channels for both the \BDT\ and the \Mbb\ analyses.
For the cross-check \Mbb\ analysis more stringent requirements are imposed on several of
the variables used for the \BDT\ selection. In addition, explicit
requirements are made on \dphiVH\
and on \Naj.
For each Higgs boson
mass, \mH, tested events are counted in a
30\GeV window centered on the mean of the expected dijet mass
peak. For the \ZllH\ modes the dijet mass distribution is asymmetric and the window is centered
5\GeV lower than \mH, while for the  \WH\ and \ZnnH\ modes
the window is centered at \mH. For these modes a
higher \pt boost requirement is made resulting in more collimated b jets
and a mass peak more symmetric around \mH. For every channel, the
\Mbb\ analysis was found to be about
10\% less sensitive than the \BDT\ analysis.

\begin{table*}[htbp]
\caption{Event selection for the \BDT\
  analysis. Where applicable, the tighter requirements for the \Mbb\
analysis are listed in parenthesis.
Entries marked ``--'' indicate that no requirement is made
for that variable.  The first two lines refer to the \pt\ threshold
on the leading ($j_1$) and sub-leading ($j_2$) jets.
CSV$_{\mathrm{max}}$ and CSV$_{\mathrm{min}}$ are the maximum and minimum b-tag requirements among the two jets.}
\label{tab:CC_BDT_sel}
\begin{center}
\begin{tabular}{cccc} \hline\hline
 Variable    & \WlnH              & \ZllH	              & \ZnnH	 \\ \hline
$\pt(j_1)$  & $>30$\GeV             & $>20$\GeV	      & $>80$\GeV	 \\
$\pt(j_2)$  & $>30$\GeV             & $>20$\GeV	      &   $>20$\GeV \\
\ptjj           & $>150$ (165)\GeV    & $>100$\GeV	      & $>160$\GeV	 \\
\ptV           & $>150$ (160)\GeV     & $>100$\GeV	      &  -- \\
\MET          & $>35$\GeV [for \WenH]  &  --	              & $>160$\GeV	 \\
\dphiVH     & -- ($>2.95$) rad      & -- ($>2.90$) rad     & -- ($>2.90$) rad	 \\
CSV$_{\mathrm{max}}$          & $>0.40$ (0.90)  & $>0.244$ (0.90)  &  $>0.50$ (0.90)	 \\
CSV$_{\mathrm{min}}$          & $>0.40$            & $>0.244$ (0.50)  & $>0.50$	 \\
\Nal           & $=0$                &  --	              & $=0$	 \\
\Naj           & -- ($=0$)            & -- ($<2$)	      &  -- ($=0$) 	 \\
\dphiMJ      & --                    &  --	              & $>0.5$ (1.5) rad	 \\ \hline
\hline
\end{tabular}
\end{center}
\end{table*}

\section{Background Control Regions}\label{sec:bck-ctrl}
Appropriate control regions that are orthogonal to the signal region are
identified in data and used to adjust the Monte Carlo simulation normalization for
the most important background processes: \Wj\ and \Zj\ (with
light- and heavy-flavor jets), and \ttbar.  For each of the search channels
and for each of these background processes, a control region is found
such that its composition is enriched in that specific background process.
The discrepancies between the expected and
observed yields in the data in these control regions are used to
obtain a scale factor by which the normalizations of
the simulations are adjusted. For each channel, this procedure is performed simultaneously for all
control regions. The background yields in the signal region from these sources are then estimated from
the adjusted simulation samples.
The uncertainties in the scale factor determination include
a statistical uncertainty due to the finite size of the samples
 and an associated systematic uncertainty from the differences in the
 shapes of the distributions that could affect the
estimate of the yields when extrapolating to the signal region. These
systematic uncertainties are obtained by varying the control
region selection criteria in order to select regions of phase space that are closer or further from the signal
region. The systematic uncertainty assigned covers the largest
variation in the scale factor value found. The procedures applied in the construction of the control regions include reversing the
b-tagging requirements to enhance \Wj\ and \Zj\ with light-flavor jets, enforcing a tighter
b-tagging requirement and requiring extra jets to enhance \ttbar, and requiring low boost in order
to enhance \Vbb\ over \ttbar.

Consistent scale factors are found for each background process
across the different channels.  For \ttbar, \Vudscg,
and \Zbb\ production the scale factors are compatible
with unity within their uncertainties ($10$--$20\%$).
For \Wbb, the control region selected contains
approximately $50\%$ \Wbb\ and single-top events, with the
remainder being \ttbar and \Wudscg, which are well
constrained by their own control regions. A choice is made to assign the observed excess of events in this region
all to \Wbb, leading to a scale factor of $2$ for this background, while the estimate
of single-top production is taken from the simulation. Reversing this
assignment has a negligible effect on the final result of the analysis.
 The total uncertainty (excluding luminosity) assigned to
the \Wbb\ yield in the signal region is approximately
$30\%$. This includes a $15\%$ uncertainty on the
extrapolation of the yield from the control region to
the signal region, determined in data with the method
outlined above. The systematic uncertainty assigned to
the predicted yield for single-top production is $30\%$.
The diboson background is taken from the simulation and a
systematic uncertainty of $30\%$ is assigned.

For \ZnnH\ the QCD multijet background in the signal region is estimated from data
using control regions of high and low values of two uncorrelated
variables with significant discriminating power towards such events.
One is the angle between the missing
energy vector and the closest jet in azimuth, \dphiMJ, and the other
is the sum of the CSV values of the two b-tagged jets. The signal
region is at high values of both discriminants, while QCD multijet events populate
regions with low values of either. The method predicts a very small
contamination of $0.015\pm 0.008$
for these background events, which is considered to be negligible.
For all other search channels,  after all selection
criteria are applied, the QCD multijet backgrounds are also found to be
negligible
and not discussed in what follows.

\section{Yield Uncertainties}\label{sec:sys}
Table~\ref{tab:syst} lists the
uncertainties on the
expected signal and background yields that enter in the limit calculation.

The uncertainty in the CMS luminosity measurement for the dataset used in the analysis is estimated to be $4.5\%$~\cite{lumiPAS}.
Muon and electron trigger,
       reconstruction, and identification efficiencies are
       determined in data from samples of
       leptonic Z boson decays. The uncertainty on the yields due to the trigger efficiency is
       2\% per charged lepton and the uncertainty on the
       identification efficiency is also 2\% per lepton. The parameters describing the
       \ZnnH\ trigger efficiency turn-on curve have been
       varied within their statistical uncertainties and for different
       assumptions on the methodology to derive the efficiency.  A yield
       uncertainty of 2\% is estimated.

The jet energy scale is
       varied within one standard deviation as a function of jet $\pt$ and
       $\eta$. The efficiency of the analysis selection is
       recomputed to assess the variation in yield.  Depending on the
       process, a 2--3\%
       yield variation is found. The effect of the uncertainty
       on the jet energy resolution is evaluated by
       smearing the jet energies according to the measured
       uncertainty. Depending on the process, a 3--6\% variation in yields due to this effect is
       obtained. An uncertainty of $3\%$ is assigned to
   the yields of all processes in the \WH\ and \ZnnH\ modes due to the
   uncertainty related to the missing transverse energy estimate.

Data-to-simulation b-tagging scale factors, measured in \ttbar events,
   are applied  consistently to jets in signal and background
       events. The measured uncertainties for the b-tagging scale factors are: $6\%$
       per b tag, $12\%$ per charm tag and of $15\%$ per
       mistagged jet (originating from gluons and light u, d, s quarks). These translate into yield uncertainties in the 3--15\%
       range, depending on the channel and the specific process.

 The total \VH\ signal cross section has
       been calculated to next-to-next-to-leading (NNLO) order accuracy,
       and the total theoretical uncertainty is $4\%$~\cite{LHCXSWG}, including
       the effect of scale and PDF variations~\cite{Botje:2011sn,Alekhin:2011sk,Lai:2010vv,Martin:2009iq,Ball:2011mu}. This analysis is performed
       in the boosted regime, and thus, potential differences in
       the \pt\ spectrum of the \V\ and \PH\ between data and Monte
       Carlo generators could introduce systematic effects in the
       signal acceptance and efficiency estimates. Calculations are available that estimate the
       next-to-leading-order (NLO)
       electroweak~\cite{HAWK1,HAWK2,HAWK3,PhysRevD.68.073003} and NNLO QCD~\cite{Grazzini,Brein}
       corrections to \VH\ production in the boosted regime. The central value used for the cross section
      in the analysis was not adjusted for these calculations. The estimated
       uncertainties from electroweak corrections for a boost of ${\sim}150\GeV$
       are $5\%$ for \cPZ\PH\ and $10\%$ for \PW\PH.  For the QCD correction, a $10\%$ uncertainty
       is estimated for both \cPZ\PH\ and \PW\PH, which includes effects due
       to additional jet activity from initial- and final-state
       radiation. The finite size of the
       signal Monte Carlo samples, after all selection criteria are applied, contributes 1--5\% uncertainty across
       all channels.

The uncertainty in the background yields that results from the
estimates from data is in the 10--35\% range.
For the predictions obtained solely from simulation, as
described in Section~\ref{sec:bck-ctrl},  an uncertainty of $30\%$  (approximately the uncertainty on the measured cross
section) is assigned  for single-top. For the diboson
backgrounds, a 30\% yield uncertainty is
assumed.

\begin{table}[htbp]
\caption{Uncertainties in the signal and background yields due to the
  uncertainty in the sources listed. The ranges
  quoted are due to variations in mode, specific process, and Higgs
  boson mass hypothesis. See text for details.}
\label{tab:syst}
\begin{center}
{\small
\begin{tabular}{cc} \hline\hline

Source                                                  &     Range   \\ \hline\hline
Luminosity                                            &      4.5\%  \\
Lepton efficiency and trigger (per lepton) &      3\%     \\
\ZnnH\ triggers                                       &      2\%    \\
Jet energy scale                                      &     2--3\% \\
Jet energy resolution                               &     3--6\%  \\
Missing transverse energy                               &      3\%    \\
b-tagging                                              &     3--15\%   \\
Signal cross section (scale and PDF)         &  4\%      \\
Signal cross section (\pt boost, EWK/QCD)    &        5--10\%/10\% \\
Signal Monte Carlo statistics                     &       1-5\% \\
Backgrounds (data estimate)                        &    10--35\%   \\
Diboson and single-top (simulation estimate)     &  30\%  \\

\hline\hline
\end{tabular}
}
\end{center}
\end{table}

\section{Results}\label{sec:results}
\begin{figure*}[htbp]
  \begin{center}
  \includegraphics[width=0.4\textwidth]{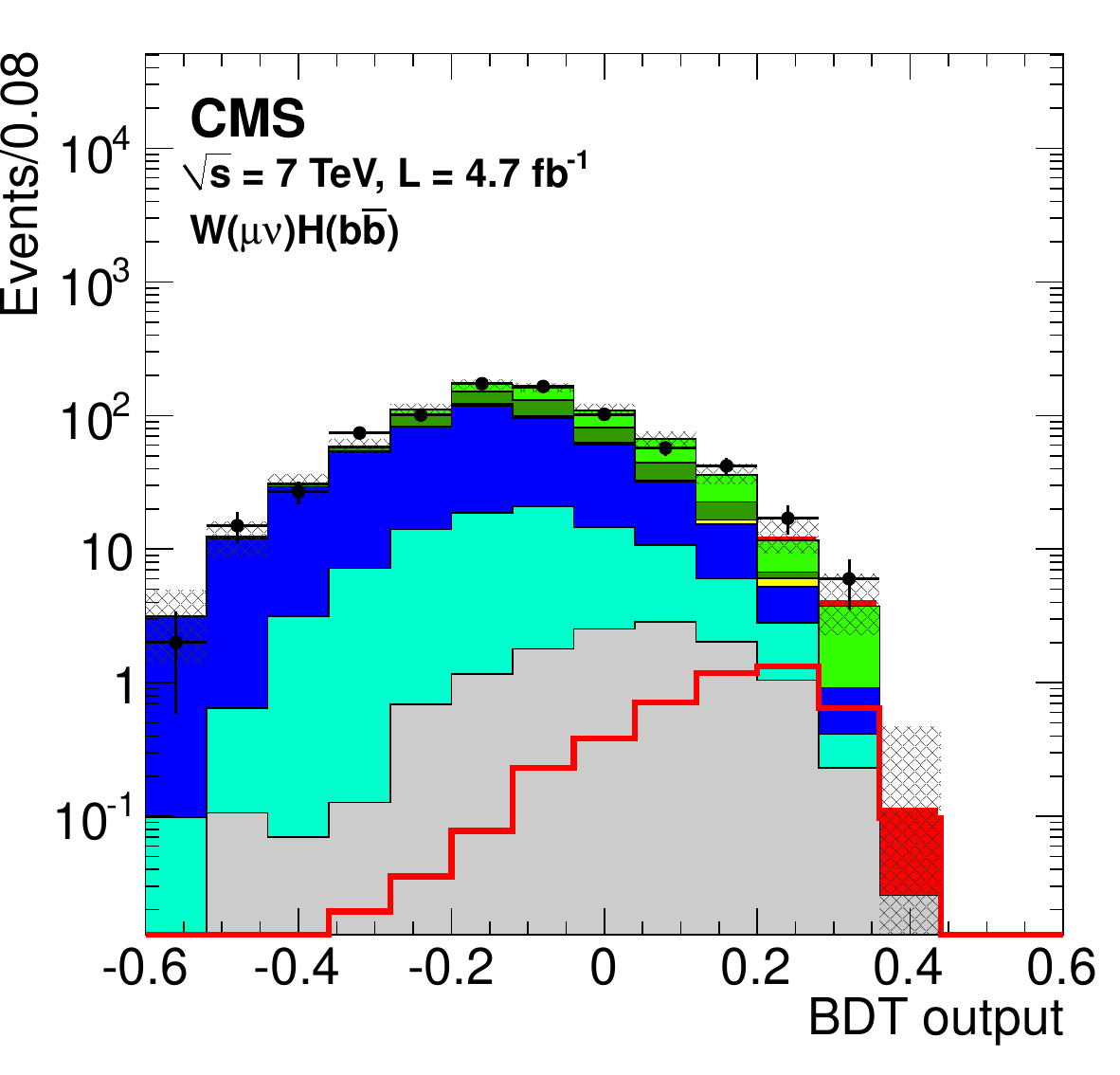}
  \includegraphics[width=0.4\textwidth]{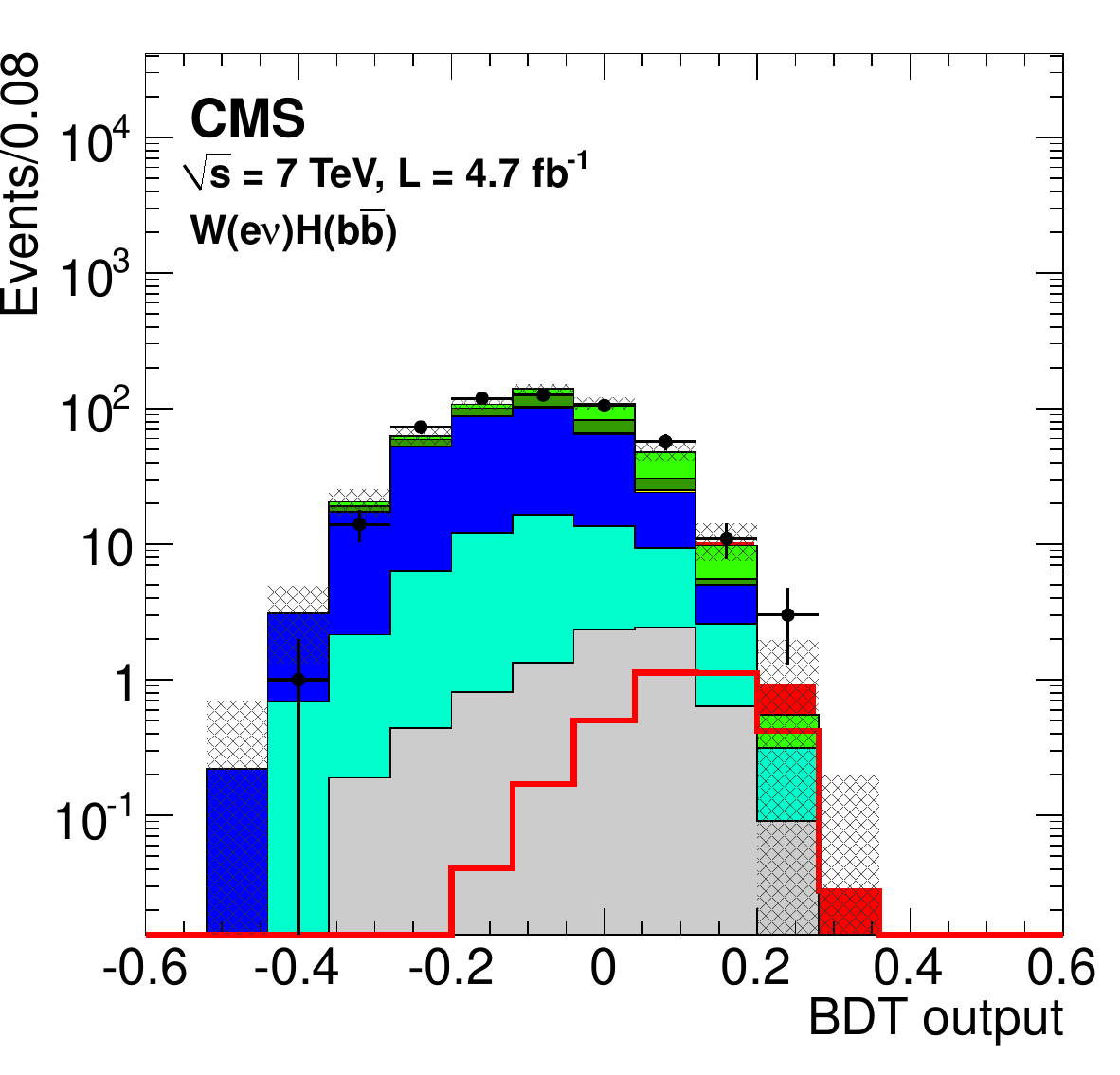}
  \includegraphics[width=0.4\textwidth]{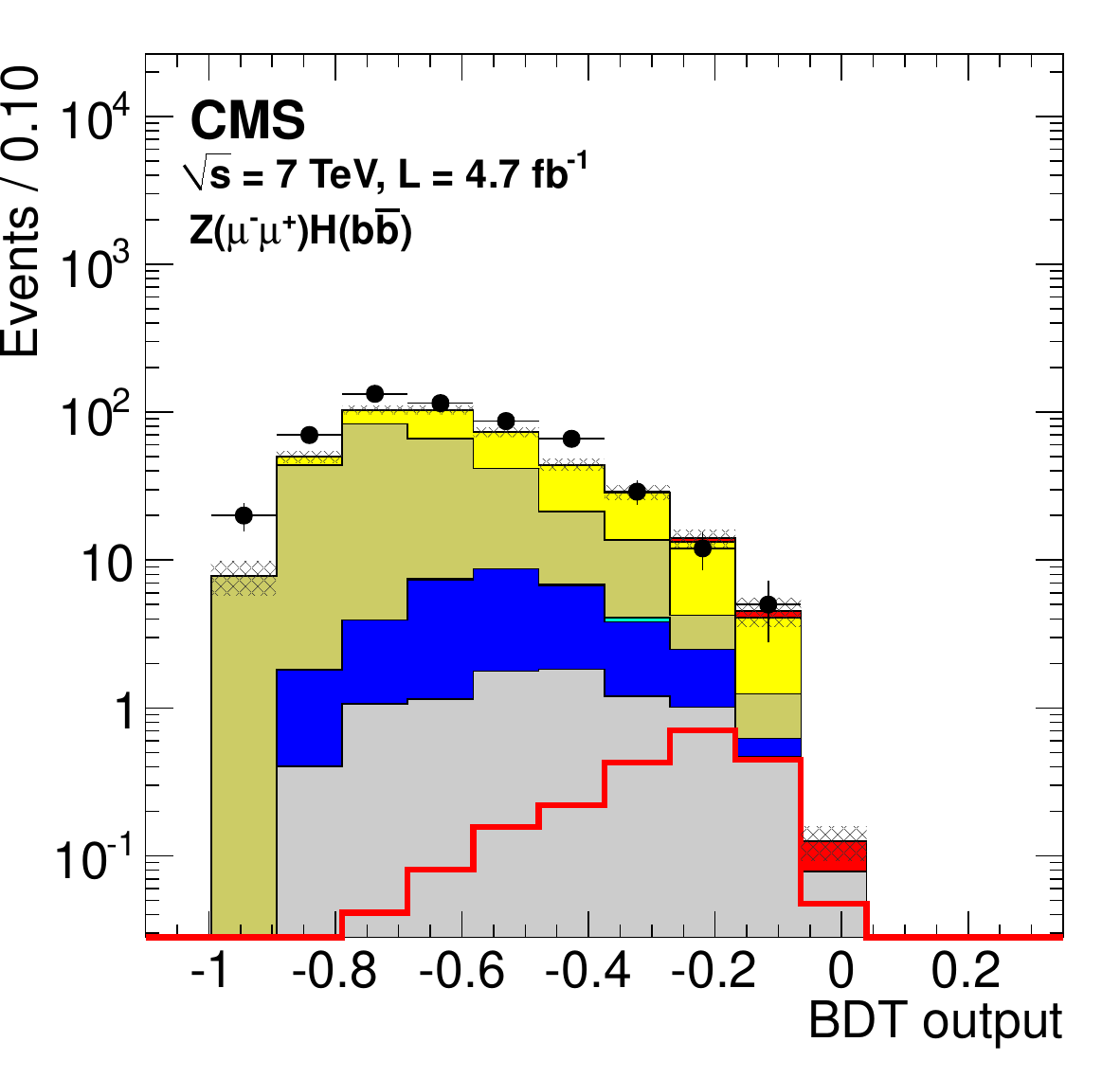}
  \includegraphics[width=0.4\textwidth]{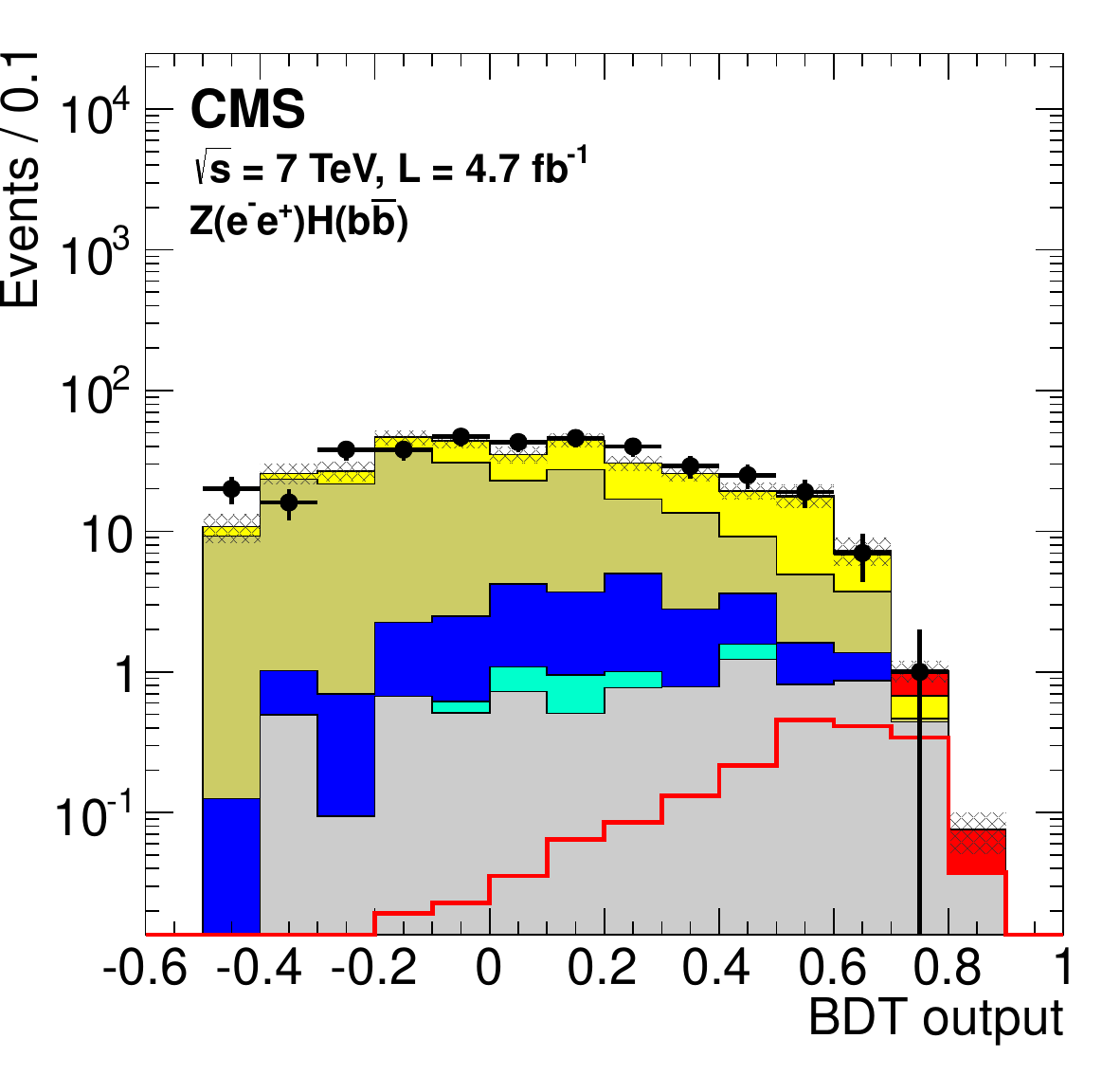}
  \includegraphics[width=0.4\textwidth]{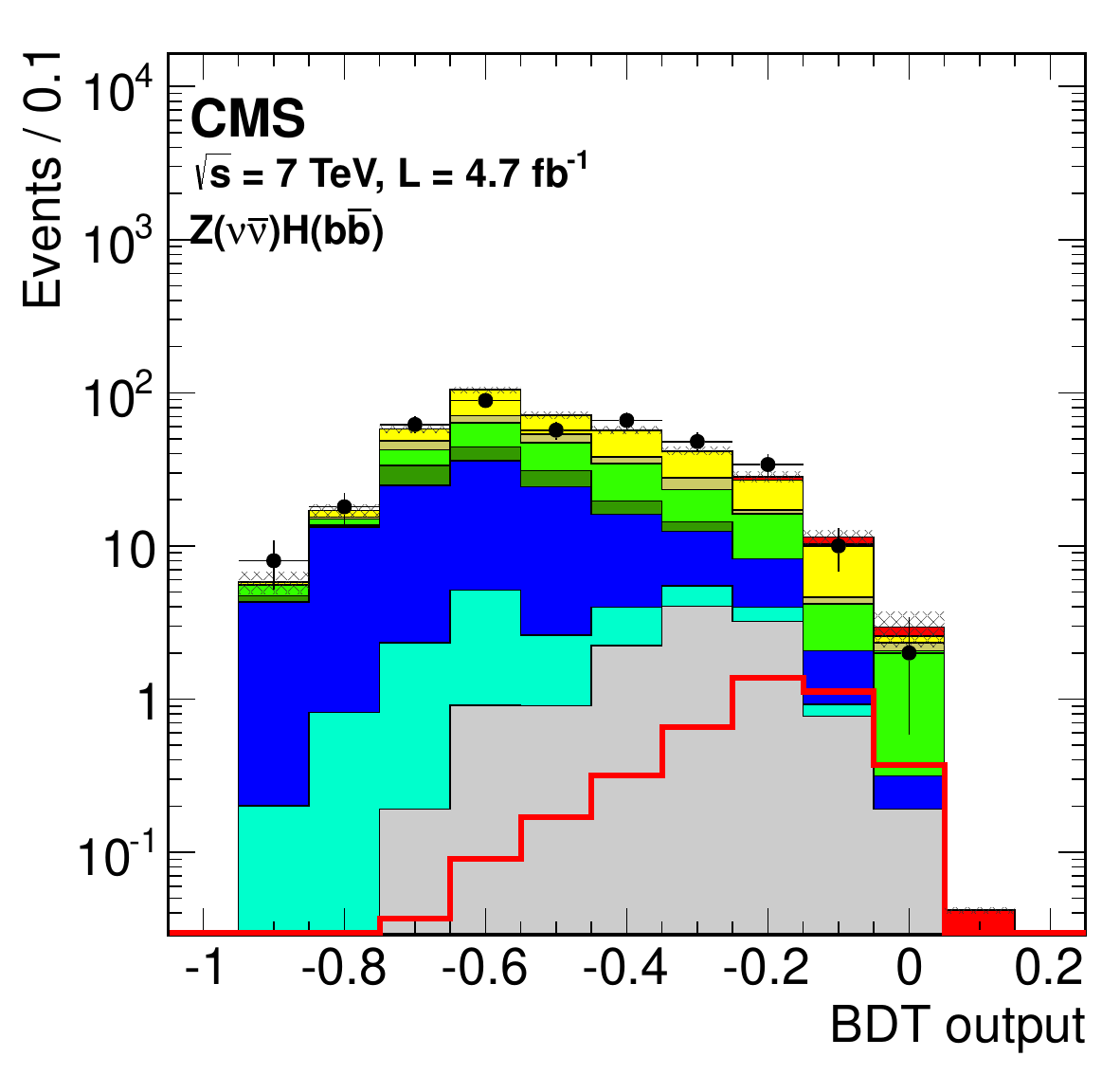}
 \includegraphics[width=0.4\textwidth]{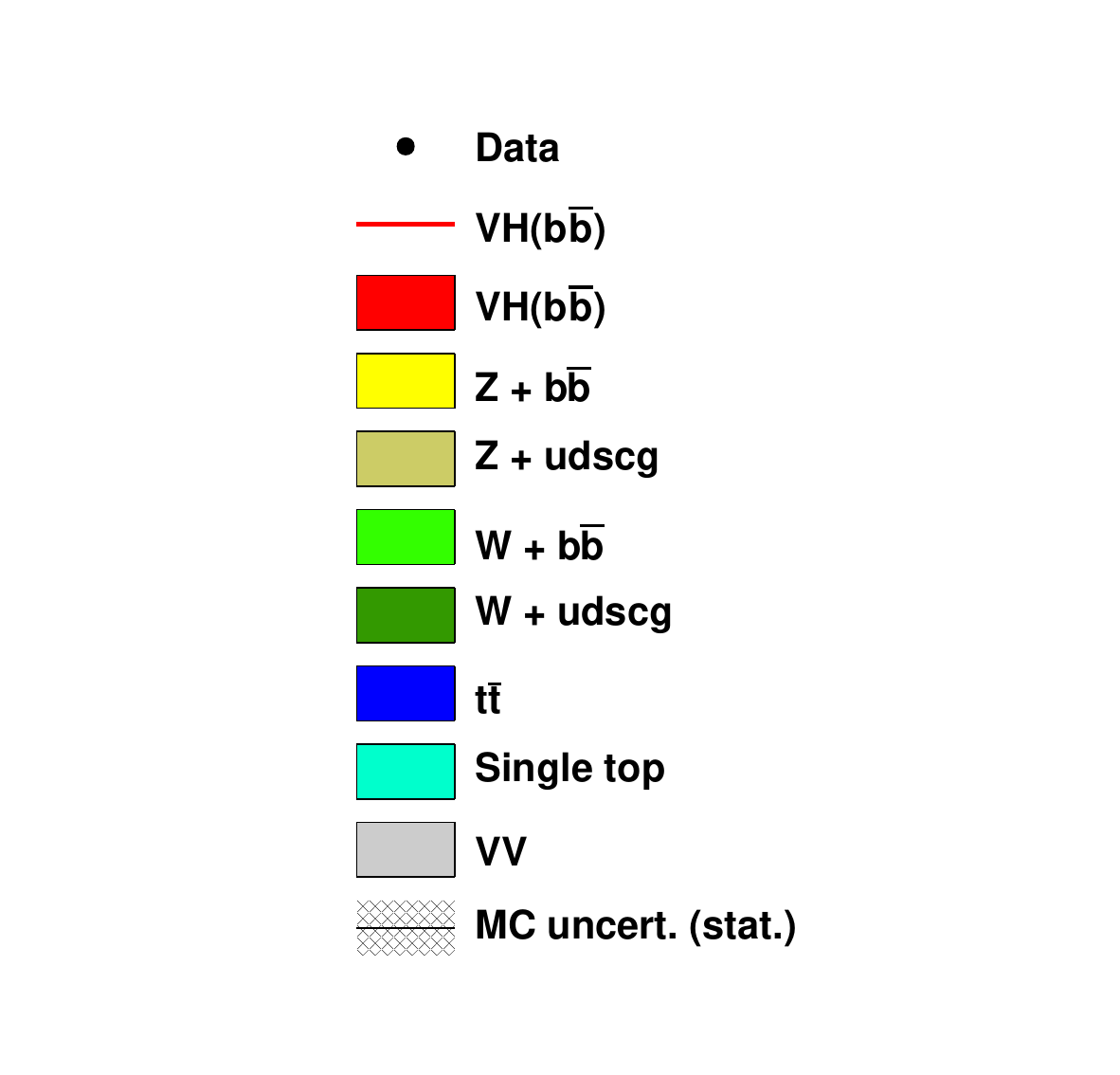}
    \caption{Distributions of the \BDT\ output, for \mH=115\GeV, for each mode after all
      selection criteria are applied. The solid histograms for the
      backgrounds and the signal are summed cumulatively. The line histogram for
      signal is also shown superimposed. The data is represented by points
      with error bars.}
    \label{fig:BDTdata}
  \end{center}
\end{figure*}

\begin{figure}[htbp]
  \begin{center}
  \includegraphics[width=0.50\textwidth]{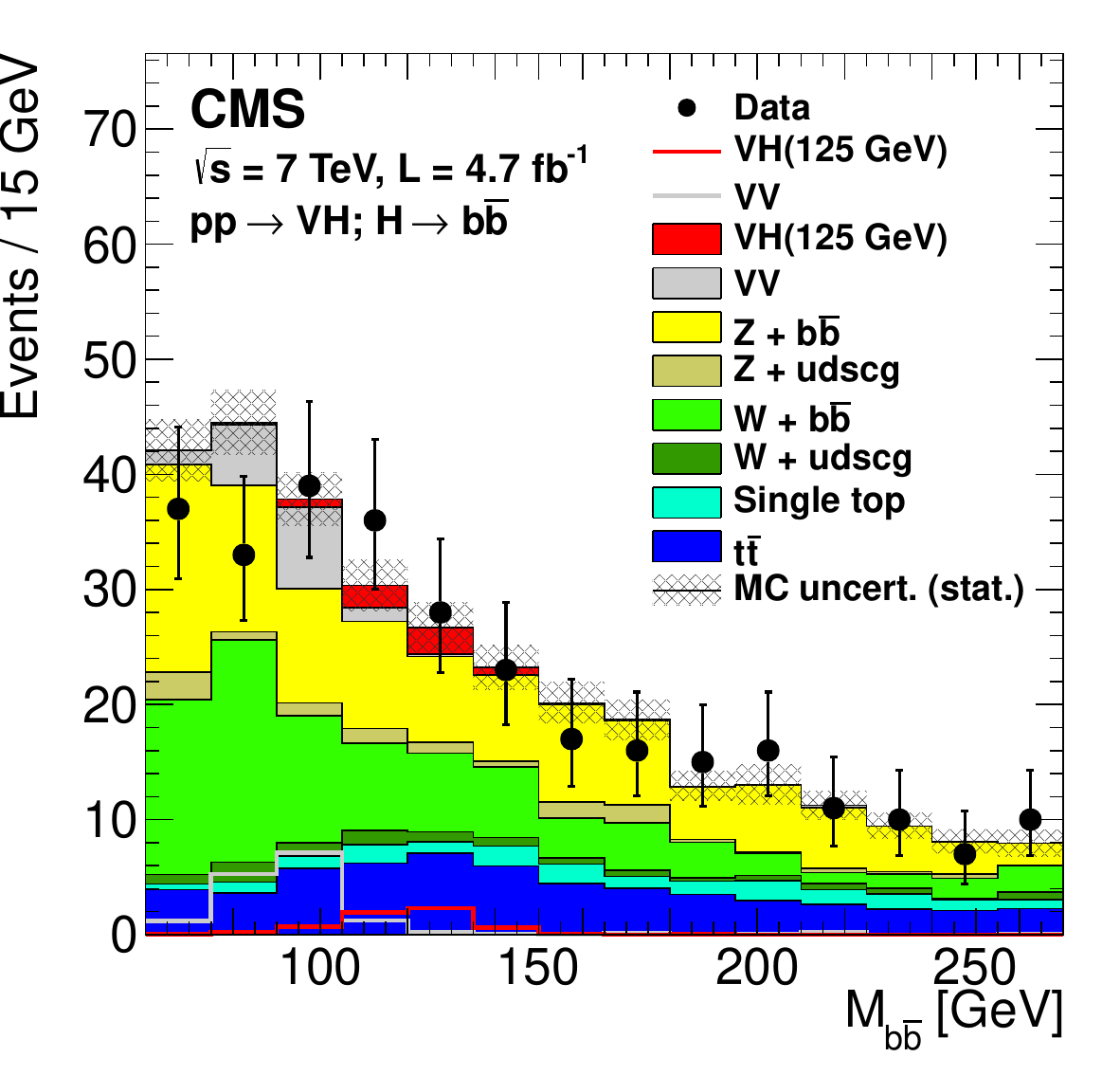}
    \caption{Dijet invariant mass distribution, combined for all
    channels, for events that pass the \Mbb\ analysis selection. The solid histograms for the
      backgrounds and the signal are summed cumulatively. The line histogram for
      signal and for \V\V\ backgrounds are also shown superimposed. The data is represented by points
      with error bars.}
    \label{fig:MJJ-combined}
  \end{center}
\end{figure}

The primary physics result presented in this Letter is an upper limit
on the production of a standard model Higgs boson in
association with a vector boson and decaying to a \bbbar\ pair.
Table~\ref{tab:BDTyields} lists, for each Higgs boson mass hypothesis considered, the expected
signal and background yields in the signal
region for the \BDT\ analysis, together with the observed number of
events. Table~\ref{tab:BDTyields} also lists the requirements on the output of the
\BDT\ distributions that define the signal region. These distributions
are shown in Fig.~\ref{fig:BDTdata} for the $\mH=115$\GeV case, where data are overlaid with the
predicted sample composition. The invariant dijet mass distribution, combined for all
    channels, for events that pass the \Mbb\ analysis selection is
    shown in Fig.~\ref{fig:MJJ-combined}.
The predicted number of background events are determined in data
using the control regions described in Section~\ref{sec:bck-ctrl},
and from direct expectations from simulation for those backgrounds for which
scale factors were not explicitly derived from control
regions. Signal yields are determined from the
simulations. The uncertainties
include all sources listed in Section~\ref{sec:sys}, except for luminosity.
Total signal uncertainties are approximately
$20$\%, and total background uncertainties are approximately in the
$20$ to $30$\%  range.

No significant excess of events is observed in any channel, and the results of all
channels are combined to obtain $95\%$ CL upper limits on the Higgs boson production
cross section in the \VH\ modes with \HBB, relative to the standard
model prediction. This is done separately for both the  \BDT\ and
\Mjj\ analyses for assumed Higgs boson masses in the 110--135\GeV
range. The observed limits at each mass point, the median expected limits and the $1\sigma$ and $2\sigma$ bands are
 calculated using the modified frequentist method
CL$_{s}$~\cite{Read1,junkcls,LHC-HCG}. The inputs
to the limit calculation include the number of observed events ($N_\text{obs}$),
and the signal and background
estimates ($B_{\rm exp}$), which are listed in
Table~\ref{tab:BDTyields} for the \BDT\ analysis.
The systematic and statistical uncertainties on the signal and background estimates, listed in
Section~\ref{sec:sys},  are treated
as nuisance parameters in the limit calculations, with
appropriate correlations taken into account.

\begin{figure}[htbp]
  \begin{center}
    \includegraphics[width=\cmsFigWidth]{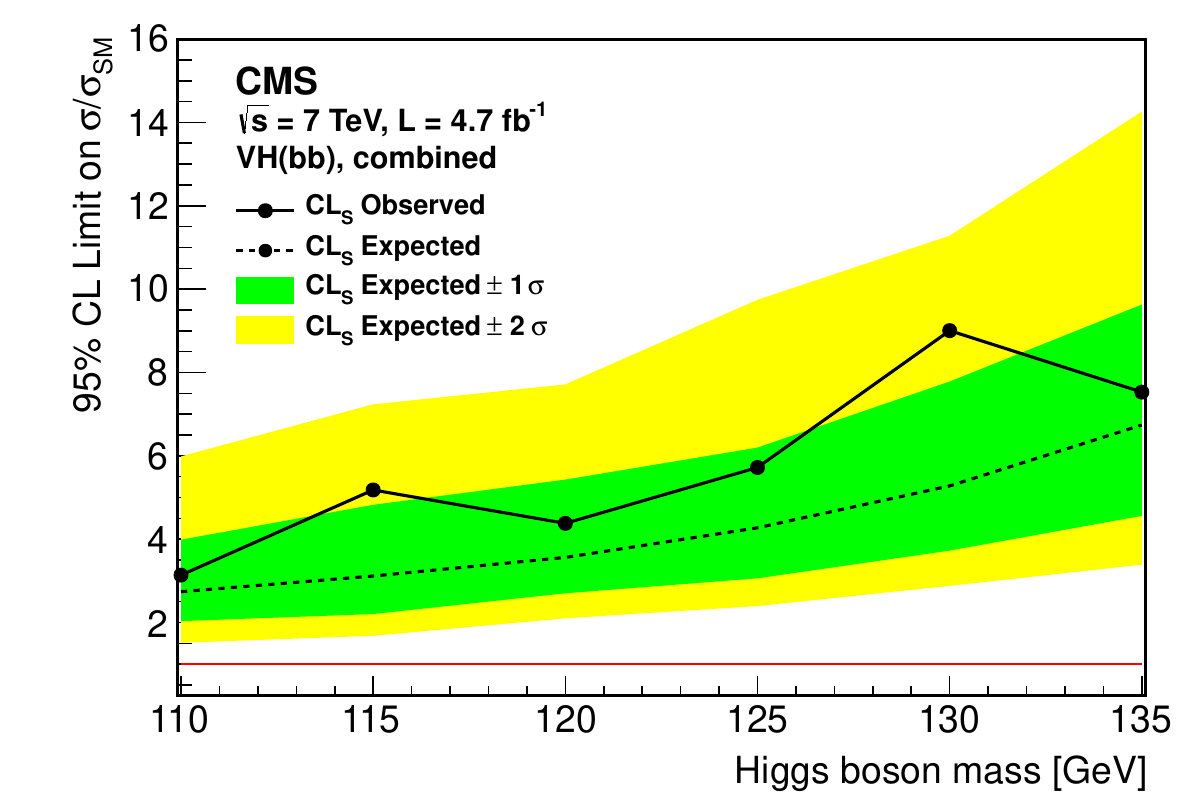}
    \caption{Expected and observed $95\%$ CL upper limits on the
product of the \VH\ production cross section times the  $\PH\to \bbbar$
branching ratio, with respect to the expectations for a standard model
Higgs boson, for the \BDT\ analysis.}
    \label{fig:Limits}
  \end{center}
\end{figure}

Table~\ref{tab:Limits} summarizes, for the \BDT\ and
\Mbb\ analyses, the expected and observed 95\% CL upper limits
on the product of the \VH\ production cross section times the  $\PH\to \bbbar$
branching ratio, with respect to the expectations for a standard model
Higgs boson ($\sigma/\sigma_{\mathrm{SM}}$).
The expected sensitivity of the \BDT\ analysis is
determined to be
superior and it is considered to be the main result in this Letter.
 The \BDT\ results are displayed
in Fig.~\ref{fig:Limits}.

\begin{table*}[htbp]
\caption{Predicted signal and background yields
 and observed number of events in data for the signal region
defined by a \BDT\ output value larger than the value listed.
The uncertainty quoted is the total uncertainty, excluding luminosity.
Results are given separately for each channel and Higgs boson mass hypothesis.
 \Wlf\ and \Zlf\ denote \Wudscg\
  and \Zudscg, respectively. ST and \V\V\ denote single top and dibosons.}
\label{tab:BDTyields}
\begin{center}
\scalebox{0.70}{

\begin{tabular}{ccccccc} \hline
\multicolumn{7}{|c|}{\WmnH}\\
\hline

Process     & $110\GeV$ & $115\GeV$ & $120\GeV$ & $125\GeV$ &
$130\GeV$ & $135\GeV$        \\\hline
\Wlf		   & $ 0.23 \pm 0.14 $  & $ 0.67 \pm 0.29 $  & $ 1.49 \pm 0.48 $  & $ 0.39 \pm 0.20 $  & $ 1.48 \pm 0.48 $  & $ 0.95 \pm 0.38 $ \\
\Wbb 	   & $ 11.04 \pm 2.55 $  & $ 7.78 \pm 1.95 $  & $ 8.32 \pm 2.04 $  & $ 4.50 \pm 1.30 $  & $ 9.01 \pm 2.16 $  & $ 6.89 \pm 1.78 $ \\
\Zbb		   & $ 0.84 \pm 0.62 $  & $ 0.84 \pm 0.62 $  & $ 1.29 \pm 0.79 $  & $ 0.84 \pm 0.62 $  & $ 1.29 \pm 0.79 $  & $ 1.29 \pm 0.79 $ \\
\ttbar		   & $ 1.66 \pm 0.59 $  & $ 2.90 \pm 0.85 $  & $ 2.90 \pm 0.84 $  & $ 1.31 \pm 0.54 $  & $ 2.79 \pm 0.80 $  & $ 1.71 \pm 0.63 $ \\
ST		   & $ 1.32 \pm 0.52 $  & $ 1.94 \pm 0.72 $  & $ 2.58 \pm 0.92 $  & $ 1.74 \pm 0.66 $  & $ 2.44 \pm 0.88 $  & $ 1.59 \pm 0.60 $ \\
\V\V		   & $ 1.93 \pm 0.66 $  & $ 1.30 \pm 0.46 $  & $ 1.12 \pm 0.40 $  & $ 0.53 \pm 0.21 $  & $ 0.76 \pm 0.29 $  & $ 0.55 \pm 0.21 $ \\
\hline
\Bexp 		   & $ 17.02 \pm 2.83 $  & $ 15.44 \pm 2.39 $  & $ 17.70 \pm 2.60 $  & $ 9.32 \pm 1.70 $  & $ 17.78 \pm 2.65 $  & $ 12.99 \pm 2.18 $ \\
\hline
Signal		   & $ 2.45 \pm 0.50 $  & $ 2.06 \pm 0.42 $  & $ 1.78 \pm 0.36 $  & $ 1.08 \pm 0.22 $  & $ 1.12 \pm 0.23 $  & $ 0.75 \pm 0.15 $ \\
\hline
\Nobs 		   & $        22       $   & $        23       $   &
$27       $   & $        15       $   & $        22       $   & $ 13
$  \\
\hline
\BDT     &  0.21  &       0.20       &    0.20      &   0.11      &    0.21 &   0.23 \\ \hline\hline
& & & & & & \\

\hline
\multicolumn{7}{|c|}{\WenH}\\
\hline

Process     & $110\GeV$ & $115\GeV$ & $120\GeV$ & $125\GeV$ &
$130\GeV$ & $135\GeV$        \\\hline
\Wlf		   & $ 0.13 \pm 0.11 $  & $ 0.51 \pm 0.26 $  & $ 0.37 \pm 0.17 $  & $ 0.23 \pm 0.12 $  & $ 0.28 \pm 0.14 $  & $ 0.10 \pm 0.11 $ \\
\Wbb 	   & $ 4.69 \pm 1.06 $  & $ 3.44 \pm 1.28 $  & $ 3.72 \pm 1.13 $  & $ 3.53 \pm 1.10 $  & $ 1.75 \pm 0.70 $  & $ 2.08 \pm 0.76 $ \\
\Zbb		   & $ 0.03 \pm 0.03 $  & $ 0.03 \pm 0.03 $  & $      -        $  & $      -        $  & $      -        $  & $      -        $ \\
\ttbar		   & $ 0.99 \pm 0.46 $  & $ 1.70 \pm 0.65 $  & $ 2.31 \pm 0.72 $  & $ 2.07 \pm 0.71 $  & $ 1.42 \pm 0.58 $  & $ 1.17 \pm 0.51 $ \\
ST		   & $ 1.59 \pm 0.59 $  & $ 1.53 \pm 0.59 $  & $ 1.75 \pm 0.67 $  & $ 1.94 \pm 1.94 $  & $ 1.51 \pm 0.59 $  & $ 1.33 \pm 0.52 $ \\
\V\V		   & $ 1.02 \pm 0.36 $  & $ 0.63 \pm 0.24 $  & $ 0.56 \pm 0.22 $  & $ 0.45 \pm 0.18 $  & $ 0.29 \pm 0.14 $  & $ 0.25 \pm 0.12 $ \\
\hline
\Bexp 		   & $ 8.46 \pm 1.36 $  & $ 7.84 \pm 1.59 $  & $ 8.72 \pm 1.52 $  & $ 8.21 \pm 2.35 $  & $ 5.25 \pm 1.10 $  & $ 4.92 \pm 1.07 $ \\
\hline
Signal		   & $ 1.63 \pm 0.34 $  & $ 1.39 \pm 0.29 $  & $ 1.20 \pm 0.25 $  & $ 1.04 \pm 0.21 $  & $ 0.76 \pm 0.16 $  & $ 0.61 \pm 0.13 $ \\
\hline
\Nobs 		   & $        9       $   & $        10       $   & $        10       $   & $        9       $   & $        8       $   & $        5       $  \\
\hline
\BDT      &  0.20  &       0.13       &    0.21      &   0.09      &    0.22 &   0.24  \\ \hline\hline
& & & & & & \\

\hline
\multicolumn{7}{|c|}{\ZmmH}\\
\hline

Process     & $110\GeV$ & $115\GeV$ & $120\GeV$ & $125\GeV$ &
$130\GeV$ & $135\GeV$        \\\hline
\Zlf		   & $ 1.16 \pm 0.59 $  & $ 0.95 \pm 0.52 $  & $ 1.67 \pm 0.72 $  & $ 0.62 \pm 0.42 $  & $ 0.81 \pm 0.48 $  & $ 1.53 \pm 0.90 $ \\
\Zbb		   & $ 4.85 \pm 1.48 $  & $ 3.14 \pm 1.06 $  & $ 7.05 \pm 1.98 $  & $ 4.38 \pm 1.48 $  & $ 5.67 \pm 1.79 $  & $ 4.06 \pm 1.52 $ \\
\ttbar		   & $ 0.64 \pm 0.22 $  & $ 0.38 \pm 0.16 $  & $ 1.05 \pm 0.32 $  & $ 0.58 \pm 0.21 $  & $ 1.19 \pm 0.32 $  & $ 0.83 \pm 0.29 $ \\
\V\V		   & $ 0.92 \pm 0.35 $  & $ 0.73 \pm 0.26 $  & $ 1.01 \pm 0.35 $  & $ 0.55 \pm 0.20 $  & $ 0.38 \pm 0.14 $  & $ 0.15 \pm 0.06 $ \\
\hline
\Bexp 		   & $ 7.57 \pm 1.64 $  & $ 5.20 \pm 1.22 $  & $ 10.78 \pm 2.16 $  & $ 6.13 \pm 1.57 $  & $ 8.05 \pm 1.88 $  & $ 6.58 \pm 1.79 $ \\
\hline
Signal		   & $ 0.92 \pm 0.17 $  & $ 0.73 \pm 0.13 $  & $ 0.88 \pm 0.16 $  & $ 0.67 \pm 0.12 $  & $ 0.59 \pm 0.11 $  & $ 0.43 \pm 0.08 $ \\
\hline
\Nobs 		   & $        7       $   & $        5       $   & $        6       $   & $        6       $   & $        11       $   & $        10       $  \\
\hline
\BDT      &    $-0.207$    &    $-0.195$   &      $-0.246$     & $-0.221$     &    $-0.313$          &          $-0.243$ \\\hline\hline
& & & & & & \\

\hline
\multicolumn{7}{|c|}{\ZeeH}\\
\hline

Process     & $110\GeV$ & $115\GeV$ & $120\GeV$ & $125\GeV$ &
$130\GeV$ & $135\GeV$        \\\hline
\Zlf		   & $ 0.02 \pm 0.02 $  & $ 0.02 \pm 0.02 $  & $ 0.20 \pm 0.19 $  & $ 0.32 \pm 0.30 $  & $ 0.36 \pm 0.35 $  & $ 0.02 \pm 0.02 $ \\
\Zbb		   & $ 2.44 \pm 0.97 $  & $ 2.51 \pm 0.98 $  & $ 5.89 \pm 2.09 $  & $ 5.48 \pm 2.04 $  & $ 2.36 \pm 0.97 $  & $ 3.44 \pm 1.19 $ \\
\ttbar		   & $ 0.11 \pm 0.08 $  & $ 0.16 \pm 0.09 $  & $ 0.38 \pm 0.17 $  & $ 0.34 \pm 0.15 $  & $      -        $  & $ 0.12 \pm 0.09 $ \\
\V\V		   & $ 1.06 \pm 0.37 $  & $ 1.07 \pm 0.38 $  & $ 1.05 \pm 0.37 $  & $ 0.92 \pm 0.33 $  & $ 0.23 \pm 0.10 $  & $ 0.46 \pm 0.19 $ \\
\hline
\Bexp 		   & $ 3.63 \pm 1.05 $  & $ 3.76 \pm 1.05 $  & $ 7.52 \pm 2.14 $  & $ 7.06 \pm 2.09 $  & $ 2.95 \pm 1.03 $  & $ 4.04 \pm 1.21 $ \\
\hline
Signal		   & $ 0.68 \pm 0.13 $  & $ 0.64 \pm 0.12 $  & $ 0.74 \pm 0.14 $  & $ 0.53 \pm 0.10 $  & $ 0.32 \pm 0.06 $  & $ 0.26 \pm 0.05 $ \\
\hline
\Nobs 		   & $        2       $   & $        4       $   & $        4       $   & $        6       $   & $        5       $   & $        4       $  \\
\hline
\BDT     & 	 $0.61$    & 	$0.63$	 & 	$0.55$     & 	 $0.59$     &    $0.65$          &          $0.67$        \\\hline\hline
& & & & & & \\

\hline
\multicolumn{7}{|c|}{\ZnnH}\\
\hline

Process     & $110\GeV$ & $115\GeV$ & $120\GeV$ & $125\GeV$ &
$130\GeV$ & $135\GeV$        \\\hline
\Wlf		   & $      -        $  & $      -        $  & $      -        $  & $ 0.89 \pm 0.18 $  & $ 1.53 \pm 0.32 $  & $ 1.53 \pm 0.32 $ \\
\Wbb 	   & $ 4.46 \pm 0.99 $  & $ 6.09 \pm 1.35 $  & $ 6.12 \pm 1.35 $  & $ 5.49 \pm 1.22 $  & $ 3.23 \pm 0.71 $  & $ 5.51 \pm 1.22 $ \\
\Zlf		   & $ 1.27 \pm 0.24 $  & $ 1.95 \pm 0.37 $  & $ 1.20 \pm 0.23 $  & $ 0.70 \pm 0.13 $  & $ 0.66 \pm 0.13 $  & $ 0.92 \pm 0.18 $ \\
\Zbb		   & $ 5.74 \pm 1.42 $  & $ 8.98 \pm 2.21 $  & $ 6.47 \pm 1.30 $  & $ 7.49 \pm 1.85 $  & $ 8.77 \pm 1.76 $  & $ 10.92 \pm 2.19 $ \\
\ttbar		   & $ 1.04 \pm 0.12 $  & $ 1.83 \pm 0.21 $  & $ 1.96 \pm 0.23 $  & $ 1.46 \pm 0.17 $  & $ 1.19 \pm 0.14 $  & $ 1.83 \pm 0.21 $ \\
ST		   & $ 0.61 \pm 0.22 $  & $ 0.85 \pm 0.31 $  & $ 0.19 \pm 0.07 $  & $ 0.27 \pm 0.10 $  & $ 0.66 \pm 0.24 $  & $ 0.53 \pm 0.19 $ \\
\V\V		   & $ 1.66 \pm 0.55 $  & $ 1.64 \pm 0.54 $  & $ 1.24 \pm 0.41 $  & $ 0.56 \pm 0.18 $  & $ 0.26 \pm 0.09 $  & $ 0.41 \pm 0.14 $ \\
\hline
\Bexp 		   & $ 14.78 \pm 1.84 $  & $ 21.34 \pm 2.70 $  & $ 17.18 \pm 1.95 $  & $ 16.86 \pm 2.24 $  & $ 16.30 \pm 1.95 $  & $ 21.65 \pm 2.55 $ \\
\hline
Signal		   & $ 1.82 \pm 0.33 $  & $ 2.23 \pm 0.40 $  & $ 1.70 \pm 0.30 $  & $ 1.64 \pm 0.29 $  & $ 1.11 \pm 0.20 $  & $ 0.98 \pm 0.18 $ \\
\hline
\Nobs 		   & $        15       $   & $        24       $   & $        20       $   & $        17       $   & $        16       $   & $        19       $  \\
\hline
\BDT  & -0.18 &  -0.17 & -0.15 & -0.20 & -0.22 &  -0.25       \\\hline\hline
& & & & & & \\

\end{tabular}

}
\end{center}
\end{table*}

\begin{table}[htbp]
\caption{Expected and observed $95\%$ CL upper limits on the
product of the \VH\ production cross section times the  $\PH\to \bbbar$
branching ratio, with respect to the expectations for a standard model
Higgs boson. The
primary results are those from the \BDT\ analysis, the \Mbb\ analysis is
presented as a cross check.}
\label{tab:Limits}
\begin{center}
{\small
\begin{tabular}{ccccccc} \hline\hline
\mH(\GeV)   &   110 & 115    & 120   & 125   & 130 & 135\\  \hline\hline
\BDT\ Exp.  &  2.7  & 3.1  & 3.6  & 4.3  & 5.3  & 6.7 \\
\BDT\ Obs.  &  3.1  & 5.2  &  4.4  &  5.7  & 9.0  & 7.5 \\ \hline
\Mbb\ Exp. & 3.0  & 3.2  & 4.4  & 4.7  & 6.4  &  7.7  \\
\Mbb\ Obs. & 3.4  & 5.6  & 6.7  & 6.3  & 10.5  & 8.9 \\

\hline\hline
\end{tabular}
}
\end{center}
\end{table}

\section{Summary}\label{sec:conclusions}

A search for the standard model Higgs boson decaying to
    \bbbar when produced in association with weak
   vector bosons
   is reported for the following channels:  $\PW(\Pgm\cPgn)\PH,$ $\PW(\Pe\cPgn)\PH,$
   $\cPZ(\Pgm\Pgm)\PH,$ $\cPZ(\Pe\Pe)\PH$ and
   $\cPZ(\nu\nu)\PH$. The search is performed in a data
   sample corresponding to an integrated luminosity of 4.7\fbinv.
   No significant excess of events above the
   expectation from background is observed. Upper limits on the
  \VH\ production cross section times the $\PH\to \bbbar$
  branching ratio, with respect to the expectations for a standard model Higgs boson, are derived for a Higgs boson in the mass range
   110--135\GeV. In this range, the observed 95\% confidence level upper limits
   vary from 3.4 to 7.5 times the standard model prediction; the
   corresponding expected limits vary from 2.7 to 6.7. This Letter reports the first upper limits from the LHC in these channels.

\section{Acknowledgements}\label{sec:ack}

We congratulate our colleagues in the CERN accelerator departments for the excellent performance of the LHC machine. We thank the technical and administrative staff at CERN and other CMS institutes, and acknowledge support from: FMSR (Austria); FNRS and FWO (Belgium); CNPq, CAPES, FAPERJ, and FAPESP (Brazil); MES (Bulgaria); CERN; CAS, MoST, and NSFC (China); COLCIENCIAS (Colombia); MSES (Croatia); RPF (Cyprus); MoER, SF0690030s09 and ERDF (Estonia); Academy of Finland, MEC, and HIP (Finland); CEA and CNRS/IN2P3 (France); BMBF, DFG, and HGF (Germany); GSRT (Greece); OTKA and NKTH (Hungary); DAE and DST (India); IPM (Iran); SFI (Ireland); INFN (Italy); NRF and WCU (Korea); LAS (Lithuania); CINVESTAV, CONACYT, SEP, and UASLP-FAI (Mexico); MSI (New Zealand); PAEC (Pakistan); MSHE and NSC (Poland); FCT (Portugal); JINR (Armenia, Belarus, Georgia, Ukraine, Uzbekistan); MON, RosAtom, RAS and RFBR (Russia); MSTD (Serbia); MICINN and CPAN (Spain); Swiss Funding Agencies (Switzerland); NSC (Taipei); TUBITAK and TAEK (Turkey); STFC (United Kingdom); DOE and NSF (USA). Individuals have received support from the Marie-Curie programme and the European Research Council (European Union); the Leventis Foundation; the A. P. Sloan Foundation; the Alexander von Humboldt Foundation; the Belgian Federal Science Policy Office; the Fonds pour la Formation \`a la Recherche dans l'Industrie et dans l'Agriculture (FRIA-Belgium); the Agentschap voor Innovatie door Wetenschap en Technologie (IWT-Belgium); the Council of Science and Industrial Research, India; and the HOMING PLUS programme of Foundation for Polish Science, cofinanced from European Union, Regional Development Fund.

\bibliography{auto_generated}   

\providecommand{\href}[2]{#2}\begingroup\raggedright\begin{thebibliography}{10}%
\makeatletter
\providecommand{\hrefCMSnoop }[0]{\@secondoftwo}%
\makeatother
\providecommand{\doi}{\texttt{doi:}\begingroup \urlstyle{tt}\Url}

\bibitem{SM1}
\hrefCMSnoop {} {{S. L. Glashow}, ``Partial Symmetries of Weak Interactions'',}
  \textit{ Nucl. Phys.} \textbf{ 22} (1961) 579,
  \href{http://dx.doi.org/10.1016/0029-5582(61)90469-2}{\doi{10.1016/0029-5582(61)90469-2}}.

\bibitem{SM2}
\hrefCMSnoop {} {{S. Weinberg}, ``A Model of Leptons'',} \textit{ Phys. Rev.
  Lett.} \textbf{ 19} (1967) 1264,
  \href{http://dx.doi.org/10.1103/PhysRevLett.19.1264}{\doi{10.1103/PhysRevLett.19.1264}}.

\bibitem{SM3}
\hrefCMSnoop {} {A.~Salam, ``Weak and electromagnetic interactions'',} in
  \textit{ Elementary particle physics: relativistic groups and analyticity},
  N.~Svartholm, ed., p.~367.
\newblock Almquvist \& Wiskell, 1968.
\newblock Proceedings of the eighth Nobel symposium.

\bibitem{Englert:1964et}
\hrefCMSnoop {} {F.~Englert and R.~Brout, ``Broken symmetry and the mass of
  gauge vector mesons'',} \textit{ Phys. Rev. Lett.} \textbf{ 13} (1964) 321,
  \href{http://dx.doi.org/10.1103/PhysRevLett.13.321}{\doi{10.1103/PhysRevLett.13.321}}.

\bibitem{Higgs:1964ia}
\hrefCMSnoop {} {P.~W. Higgs, ``Broken symmetries, massless particles and gauge
  fields'',} \textit{ Phys. Lett.} \textbf{ 12} (1964) 132,
  \href{http://dx.doi.org/10.1016/0031-9163(64)91136-9}{\doi{10.1016/0031-9163(64)91136-9}}.

\bibitem{Higgs:1964pj}
\hrefCMSnoop {} {P.~W. Higgs, ``Broken symmetries and the masses of gauge
  bosons'',} \textit{ Phys. Rev. Lett.} \textbf{ 13} (1964) 508,
  \href{http://dx.doi.org/10.1103/PhysRevLett.13.508}{\doi{10.1103/PhysRevLett.13.508}}.

\bibitem{Guralnik:1964eu}
\hrefCMSnoop {} {G.~S. Guralnik, C.~R. Hagen, and T.~W.~B. Kibble, ``Global
  conservation laws and massless particles'',} \textit{ Phys. Rev. Lett.}
  \textbf{ 13} (1964) 585,
  \href{http://dx.doi.org/10.1103/PhysRevLett.13.585}{\doi{10.1103/PhysRevLett.13.585}}.

\bibitem{Higgs:1966ev}
\hrefCMSnoop {} {P.~W. Higgs, ``Spontaneous symmetry breakdown without massless
  bosons'',} \textit{ Phys. Rev.} \textbf{ 145} (1966) 1156,
  \href{http://dx.doi.org/10.1103/PhysRev.145.1156}{\doi{10.1103/PhysRev.145.1156}}.

\bibitem{Kibble:1967sv}
\hrefCMSnoop {} {T.~W.~B. Kibble, ``Symmetry breaking in non-{A}belian gauge
  theories'',} \textit{ Phys. Rev.} \textbf{ 155} (1967) 1554,
  \href{http://dx.doi.org/10.1103/PhysRev.155.1554}{\doi{10.1103/PhysRev.155.1554}}.

\bibitem{LEPHIGGS}
\hrefCMSnoop {} {{ALEPH, DELPHI, L3, OPAL Collaborations, and the LEP Working
  Group for {H}iggs Boson Searches}, ``Search for the Standard Model {H}iggs
  boson at {LEP}'',} \textit{ Phys. Lett. B} \textbf{ 565} (2003) 61,
  \href{http://dx.doi.org/10.1016/S0370-2693(03)00614-2}{\doi{10.1016/S0370-2693(03)00614-2}}.

\bibitem{TEVHIGGS_2010}
\hrefCMSnoop {} {{CDF and D0 Collaborations}, ``Combination of {T}evatron
  Searches for the Standard Model {H}iggs Boson in the {$\WW$} Decay Mode'',}
  \textit{ Phys. Rev. Lett.} \textbf{ 104} (2010) 061802,
  \href{http://dx.doi.org/10.1103/PhysRevLett.104.061802}{\doi{10.1103/PhysRevLett.104.061802}}.
  {A more recent, unpublished, limit is given in preprint arXiv:1103.3233.}

\bibitem{ATLAS:2011aa}
\hrefCMSnoop {} {{ ATLAS} Collaboration, ``{Search for the Higgs boson in the
  H$\to$WW(*)$\to \ell^+ \nu \ell^- \bar{\nu}$ decay channel in pp collisions
  at $\sqrt{s}$ = 7 TeV with the ATLAS detector}'',} (2011).
  \href{http://www.arXiv.org/abs/1112.2577}{\texttt{ arXiv:1112.2577}}.
Submitted to \textit{Phys. Rev. Rev. Lett.}

\bibitem{Aad:2011uq}
\hrefCMSnoop {} {{ ATLAS} Collaboration, ``{Search for the standard model Higgs
  boson in the decay channel $H \to ZZ^{(*)} \to 4\ell$ with the ATLAS
  detector}'',} \textit{ Phys. Lett. B} \textbf{ 705} (2011) 435,
  \href{http://dx.doi.org/http://dx.doi.org/10.1016/j.physletb.2011.10.034}{\doi{http://dx.doi.org/10.1016/j.physletb.2011.10.034}},
\href{http://www.arXiv.org/abs/1109.5945}{\texttt{ arXiv:1109.5945}}.

\bibitem{ATLAS:2011af}
\hrefCMSnoop {} {{ ATLAS} Collaboration, ``{Search for a standard model Higgs
  boson in the $H \to ZZ \to \ell^+\ell^-\nu \bar{\nu}$ decay channel with the
  ATLAS detector}'',} \textit{ Phys. Rev. Lett.} \textbf{ 107} (2011) 221802,
  \href{http://dx.doi.org/10.1103/PhysRevLett.107.221802}{\doi{10.1103/PhysRevLett.107.221802}},
\href{http://www.arXiv.org/abs/1109.3357}{\texttt{ arXiv:1109.3357}}.

\bibitem{LEPewkfits}
\hrefCMSnoop {} {{ALEPH, CDF, D0, DELPHI, L3, OPAL, SLD Collaborations, the LEP
  Working Group, the Tevatron Electroweak Working Group, and the SLD
  Electroweak and Heavy flavor Group}, ``Precision electroweak measurements and
  constraints on the standard model'',} (2010).
\href{http://www.arXiv.org/abs/1012.2367}{\texttt{ arXiv:1012.2367}}.

\bibitem{Aaltonen:2009dh}
\hrefCMSnoop {} {{ CDF} Collaboration, ``Search for a Standard Model {H}iggs
  Boson produced in {$WH\to\ell b\overline{b}$} in $p\overline{p}$ Collisions
  at $\sqrt{s}=1.96$ {TeV}'',} \textit{ Phys. Rev. Lett.} \textbf{ 103} (2009)
  101802,
  \href{http://dx.doi.org/10.1103/PhysRevLett.103.101802}{\doi{10.1103/PhysRevLett.103.101802}},
\href{http://www.arXiv.org/abs/0906.5613}{\texttt{ arXiv:0906.5613}}.

\bibitem{Aaltonen:2009jg}
\hrefCMSnoop {} {{ CDF} Collaboration, ``A Search for the {H}iggs Boson Using
  Neural Networks in Events with Missing Energy and b-quark Jets in
  $p\overline{p}$ Collisions at $\sqrt{s} = 1.96$ {TeV}'',} \textit{ Phys. Rev.
  Lett.} \textbf{ 104} (2010) 141801,
  \href{http://dx.doi.org/10.1103/PhysRevLett.104.141801}{\doi{10.1103/PhysRevLett.104.141801}},
\href{http://www.arXiv.org/abs/0911.3935}{\texttt{ arXiv:0911.3935}}.

\bibitem{Aaltonen:2010pa}
\hrefCMSnoop {} {{ CDF} Collaboration, ``Improved Search for a {H}iggs Boson
  Produced in Association with {$Z\rightarrow{}{l}^{+}{l}^{-}$ in
  $p\overline{p}$} Collisions at
  {$\sqrt{s}=1.96\text{\,}\text{\,}\mathrm{TeV}$}'',} \textit{ Phys. Rev.
  Lett.} \textbf{ 105} (2010) 251802,
  \href{http://dx.doi.org/10.1103/PhysRevLett.105.251802}{\doi{10.1103/PhysRevLett.105.251802}},
\href{http://www.arXiv.org/abs/1009.3047}{\texttt{ arXiv:1009.3047}}.

\bibitem{Aaltonen:2011dh}
\hrefCMSnoop {} {{ CDF} Collaboration, ``{Search for the Higgs boson in the
  all-hadronic final state using the CDF II detector}'',} \textit{ Phys. Rev.
  D} \textbf{ 84} (2011) 052010,
  \href{http://dx.doi.org/10.1103/PhysRevD.84.052010}{\doi{10.1103/PhysRevD.84.052010}},
  \href{http://www.arXiv.org/abs/1102.0024}{\texttt{ arXiv:1102.0024}}.
submitted to Phys. Rev. D.

\bibitem{Abazov20116}
\hrefCMSnoop {} {{ D0} Collaboration, ``Search for {WH} associated production
  in 5.3 fb$^{-1}$ of collisions at the {F}ermilab {T}evatron'',} \textit{
  Phys. Lett. B} \textbf{ 698} (2011), no.~1, 6,
  \href{http://dx.doi.org/10.1016/j.physletb.2011.02.036}{\doi{10.1016/j.physletb.2011.02.036}}.

\bibitem{AbazovZnnHbb}
\hrefCMSnoop {} {{ D0} Collaboration, ``Search for the Standard Model {H}iggs
  Boson in the {$ZH\rightarrow{}\nu{}\overline{\nu{}}b\overline{b}$} Channel in
  $5.2\text{\,}\text{\,}{\mathrm{fb}}^{-1}$ of $p\overline{p}$ Collisions at
  {$\sqrt{s}=1.96\text{\,}\text{\,}\mathrm{TeV}$}'',} \textit{ Phys. Rev.
  Lett.} \textbf{ 104} (Feb, 2010) 071801,
  \href{http://dx.doi.org/10.1103/PhysRevLett.104.071801}{\doi{10.1103/PhysRevLett.104.071801}}.

\bibitem{AbazovZllHbb}
\hrefCMSnoop {} {{ D0} Collaboration, ``Search for
  {$ZH\rightarrow{}{l}^{+}{l}^{-}b\overline{b}$} Production in
  $4.2\text{\,}\text{\,}{\mathrm{fb}}^{-1}$ of $p\overline{p}$ Collisions at
  {$\sqrt{s}=1.96\text{\,}\text{\,}\mathrm{TeV}$}'',} \textit{ Phys. Rev.
  Lett.} \textbf{ 105} (Dec, 2010) 251801,
  \href{http://dx.doi.org/10.1103/PhysRevLett.105.251801}{\doi{10.1103/PhysRevLett.105.251801}}.

\bibitem{Ellis:1975ap}
\hrefCMSnoop {} {J.~R. Ellis, M.~K. Gaillard, and D.~V. Nanopoulos, ``A
  phenomenological profile of the {H}iggs boson'',} \textit{ Nucl. Phys. B}
  \textbf{ 106} (1976) 292,
\href{http://dx.doi.org/10.1016/0550-3213(76)90382-5}{\doi{10.1016/0550-3213(76)90382-5}}.

\bibitem{Georgi:1977gs}
\hrefCMSnoop {} {H.~Georgi, S.~Glashow, M.~Machacek, and D.~V. Nanopoulos,
  ``{H}iggs Bosons from Two Gluon Annihilation in Proton Proton Collisions'',}
  \textit{ Phys. Rev. Lett.} \textbf{ 40} (1978) 692,
\href{http://dx.doi.org/10.1103/PhysRevLett.40.692}{\doi{10.1103/PhysRevLett.40.692}}.

\bibitem{Djouadi:1991tka}
\hrefCMSnoop {} {A.~Djouadi, M.~Spira, and P.~M. Zerwas, ``{Production of Higgs
  bosons in proton colliders: QCD corrections}'',} \textit{ Phys. Lett. B}
  \textbf{ 264} (1991) 440,
\href{http://dx.doi.org/10.1016/0370-2693(91)90375-Z}{\doi{10.1016/0370-2693(91)90375-Z}}.

\bibitem{Dawson:1990zj}
\hrefCMSnoop {} {S.~Dawson, ``{Radiative corrections to Higgs boson
  production}'',} \textit{ Nucl. Phys. B} \textbf{ 359} (1991) 283,
\href{http://dx.doi.org/10.1016/0550-3213(91)90061-2}{\doi{10.1016/0550-3213(91)90061-2}}.

\bibitem{Spira:1995rr}
\hrefCMSnoop {} {M.~Spira {et~al.}, ``{Higgs boson production at the LHC}'',}
  \textit{ Nucl. Phys. B} \textbf{ 453} (1995) 17,
  \href{http://dx.doi.org/10.1016/0550-3213(95)00379-7}{\doi{10.1016/0550-3213(95)00379-7}},
\href{http://www.arXiv.org/abs/hep-ph/9504378}{\texttt{ arXiv:hep-ph/9504378}}.

\bibitem{Harlander:2002wh}
\hrefCMSnoop {} {R.~V. Harlander and W.~B. Kilgore, ``{Next-to-next-to-leading
  order Higgs production at hadron colliders}'',} \textit{ Phys. Rev. Lett.}
  \textbf{ 88} (2002) 201801,
  \href{http://dx.doi.org/10.1103/PhysRevLett.88.201801}{\doi{10.1103/PhysRevLett.88.201801}},
\href{http://www.arXiv.org/abs/hep-ph/0201206}{\texttt{ arXiv:hep-ph/0201206}}.

\bibitem{Anastasiou:2002yz}
\hrefCMSnoop {} {C.~Anastasiou and K.~Melnikov, ``{Higgs boson production at
  hadron colliders in NNLO QCD}'',} \textit{ Nucl. Phys. B} \textbf{ 646}
  (2002) 220,
  \href{http://dx.doi.org/10.1016/S0550-3213(02)00837-4}{\doi{10.1016/S0550-3213(02)00837-4}},
\href{http://www.arXiv.org/abs/hep-ph/0207004}{\texttt{ arXiv:hep-ph/0207004}}.

\bibitem{Ravindran:2003um}
\hrefCMSnoop {} {V.~Ravindran, J.~Smith, and W.~L. van Neerven, ``{NNLO
  corrections to the total cross section for Higgs boson production in hadron
  hadron collisions}'',} \textit{ Nucl. Phys. B} \textbf{ 665} (2003) 325,
  \href{http://dx.doi.org/10.1016/S0550-3213(03)00457-7}{\doi{10.1016/S0550-3213(03)00457-7}},
\href{http://www.arXiv.org/abs/hep-ph/0302135}{\texttt{ arXiv:hep-ph/0302135}}.

\bibitem{Catani:2003zt}
\hrefCMSnoop {} {S.~Catani {et~al.}, ``{Soft-gluon resummation for Higgs boson
  production at hadron colliders}'',} \textit{ JHEP} \textbf{ 07} (2003) 028,
  \href{http://dx.doi.org/10.1088/1126-6708/2003/07/028}{\doi{10.1088/1126-6708/2003/07/028}},
\href{http://www.arXiv.org/abs/hep-ph/0306211}{\texttt{ arXiv:hep-ph/0306211}}.

\bibitem{Aglietti:2004nj}
\hrefCMSnoop {} {U.~Aglietti, R.~Bonciani, G.~Degrassi, and A.~Vicini,
  ``{Two-loop light fermion contribution to Higgs production and decays}'',}
  \textit{ Phys. Lett. B} \textbf{ 595} (2004) 432,
  \href{http://dx.doi.org/10.1016/j.physletb.2004.06.063}{\doi{10.1016/j.physletb.2004.06.063}},
\href{http://www.arXiv.org/abs/hep-ph/0404071}{\texttt{ arXiv:hep-ph/0404071}}.

\bibitem{Degrassi:2004mx}
\hrefCMSnoop {} {G.~Degrassi and F.~Maltoni, ``{Two-loop electroweak
  corrections to Higgs production at hadron colliders}'',} \textit{ Phys. Lett.
  B} \textbf{ 600} (2004) 255,
  \href{http://dx.doi.org/10.1016/j.physletb.2004.09.008}{\doi{10.1016/j.physletb.2004.09.008}},
\href{http://www.arXiv.org/abs/hep-ph/0407249}{\texttt{ arXiv:hep-ph/0407249}}.

\bibitem{Baglio:2010ae}
\hrefCMSnoop {} {J.~Baglio and A.~Djouadi, ``{Higgs production at the LHC}'',}
  \textit{ JHEP} \textbf{ 03} (2011) 055,
  \href{http://dx.doi.org/10.1007/JHEP03(2011)055}{\doi{10.1007/JHEP03(2011)055}},
\href{http://www.arXiv.org/abs/1012.0530}{\texttt{ arXiv:1012.0530}}.

\bibitem{Actis:2008ug}
\hrefCMSnoop {} {S.~Actis {et~al.}, ``{NLO} electroweak corrections to {H}iggs
  boson production at hadron colliders'',} \textit{ Phys. Lett. B} \textbf{
  670} (2008) 12,
  \href{http://dx.doi.org/10.1016/j.physletb.2008.10.018}{\doi{10.1016/j.physletb.2008.10.018}},
\href{http://www.arXiv.org/abs/0809.1301}{\texttt{ arXiv:0809.1301}}.

\bibitem{Anastasiou:2008tj}
\hrefCMSnoop {} {C.~Anastasiou, R.~Boughezal, and F.~Petriello, ``{Mixed
  QCD-electroweak corrections to Higgs boson production in gluon fusion}'',}
  \textit{ JHEP} \textbf{ 04} (2009) 003,
  \href{http://dx.doi.org/10.1088/1126-6708/2009/04/003}{\doi{10.1088/1126-6708/2009/04/003}},
\href{http://www.arXiv.org/abs/0811.3458}{\texttt{ arXiv:0811.3458}}.

\bibitem{deFlorian:2009hc}
\hrefCMSnoop {} {D.~de~Florian and M.~Grazzini, ``{Higgs production through
  gluon fusion: updated cross sections at the Tevatron and the LHC}'',}
  \textit{ Phys. Lett. B} \textbf{ 674} (2009) 291,
  \href{http://dx.doi.org/10.1016/j.physletb.2009.03.033}{\doi{10.1016/j.physletb.2009.03.033}},
\href{http://www.arXiv.org/abs/0901.2427}{\texttt{ arXiv:0901.2427}}.

\bibitem{Djouadi:1997yw}
\hrefCMSnoop {} {A.~Djouadi, J.~Kalinowski, and M.~Spira, ``{HDECAY: A program
  for Higgs boson decays in the standard model and its supersymmetric
  extension}'',} \textit{ Comput. Phys. Commun.} \textbf{ 108} (1998) 56,
  \href{http://dx.doi.org/10.1016/S0010-4655(97)00123-9}{\doi{10.1016/S0010-4655(97)00123-9}},
\href{http://www.arXiv.org/abs/hep-ph/9704448}{\texttt{ arXiv:hep-ph/9704448}}.

\bibitem{LHCXSWG}
\href {http://cdsweb.cern.ch/record/1318996} {{ {LHC Higgs Cross Section
  Working Group}} Collaboration, ``Handbook of {LHC} {H}iggs cross sections: 1.
  Inclusive Observables'',} CERN Report CERN-2011-002, 2011.
\newblock \href{http://www.arXiv.org/abs/1101.0593}{\texttt{ arXiv:1101.0593}}.

\bibitem{Ciccolini:2007jr}
\hrefCMSnoop {} {M.~Ciccolini, A.~Denner, and S.~Dittmaier, ``{Strong and
  electroweak corrections to the production of Higgs + 2-jets via weak
  interactions at the LHC}'',} \textit{ Phys. Rev. Lett.} \textbf{ 99} (2007)
  161803,
  \href{http://dx.doi.org/10.1103/PhysRevLett.99.161803}{\doi{10.1103/PhysRevLett.99.161803}},
  \href{http://www.arXiv.org/abs/0707.0381}{\texttt{ arXiv:0707.0381}}.

\bibitem{Ciccolini:2007ec}
\hrefCMSnoop {} {M.~Ciccolini, A.~Denner, and S.~Dittmaier, ``{Electroweak and
  QCD corrections to Higgs production via vector-boson fusion at the LHC}'',}
  \textit{ Phys. Rev. D} \textbf{ 77} (2008) 013002,
  \href{http://dx.doi.org/10.1103/PhysRevD.77.013002}{\doi{10.1103/PhysRevD.77.013002}},
\href{http://www.arXiv.org/abs/0710.4749}{\texttt{ arXiv:0710.4749}}.

\bibitem{Figy:2003nv}
\hrefCMSnoop {} {T.~Figy, C.~Oleari, and D.~Zeppenfeld, ``{Next-to-leading
  order jet distributions for Higgs boson production via weak-boson fusion}'',}
  \textit{ Phys. Rev. D} \textbf{ 68} (2003) 073005,
  \href{http://dx.doi.org/10.1103/PhysRevD.68.073005}{\doi{10.1103/PhysRevD.68.073005}},
\href{http://www.arXiv.org/abs/hep-ph/0306109}{\texttt{ arXiv:hep-ph/0306109}}.

\bibitem{Arnold:2008rz}
\hrefCMSnoop {} {K.~Arnold {et~al.}, ``{VBFNLO: A parton level Monte Carlo for
  processes with electroweak bosons}'',} \textit{ Comput. Phys. Commun.}
  \textbf{ 180} (2009) 1661,
  \href{http://dx.doi.org/10.1016/j.cpc.2009.03.006}{\doi{10.1016/j.cpc.2009.03.006}},
\href{http://www.arXiv.org/abs/0811.4559}{\texttt{ arXiv:0811.4559}}.

\bibitem{Bolzoni:2010xr}
\hrefCMSnoop {} {P.~Bolzoni {et~al.}, ``{Higgs production via vector-boson
  fusion at NNLO in QCD}'',} \textit{ Phys. Rev. Lett.} \textbf{ 105} (2010)
  011801,
  \href{http://dx.doi.org/10.1103/PhysRevLett.105.011801}{\doi{10.1103/PhysRevLett.105.011801}},
\href{http://www.arXiv.org/abs/1003.4451}{\texttt{ arXiv:1003.4451}}.

\bibitem{PhysRevD.18.1724}
\hrefCMSnoop {} {S.~L. Glashow, D.~V. Nanopoulos, and A.~Yildiz, ``Associated
  production of {H}iggs bosons and {Z} particles'',} \textit{ Phys. Rev. D}
  \textbf{ 18} (Sep, 1978) 1724,
  \href{http://dx.doi.org/10.1103/PhysRevD.18.1724}{\doi{10.1103/PhysRevD.18.1724}}.

\bibitem{Roe:2005hm}
\href
  {http://www.physics.ox.ac.uk/phystat05/proceedings/files/phystat05-proc.pdf}
  {B.~P. Roe, H.-J. Yang, and J.~Zhu, ``Boosted decision trees, a powerful
  event classifier'',} in \textit{ Proceedings of PHYSTATO5: Statistical
  Problems in Particle Physics, Astrophysics and Cosmology}.
\newblock 2005.

\bibitem{CMSDETECTOR}
\hrefCMSnoop {} {{ CMS} Collaboration, ``The CMS experiment at the CERN LHC'',}
  \textit{ JINST} \textbf{ 03} (2008) S08004,
  \href{http://dx.doi.org/10.1088/1748-0221/3/08/S08004}{\doi{10.1088/1748-0221/3/08/S08004}}.

\bibitem{GEANT4}
\hrefCMSnoop {} {{ GEANT4} Collaboration, ``{GEANT4}--a simulation toolkit'',}
  \textit{ Nucl. Instrum. Meth. A} \textbf{ 506} (2003) 250,
  \href{http://dx.doi.org/10.1016/S0168-9002(03)01368-8}{\doi{10.1016/S0168-9002(03)01368-8}}.

\bibitem{POWHEG}
\hrefCMSnoop {} {S.~Frixione, P.~Nason, and C.~Oleari, ``Matching {NLO} {QCD}
  computations with Parton Shower simulations: the {POWHEG} method'',} \textit{
  JHEP} \textbf{ 11} (2007) 070,
  \href{http://dx.doi.org/10.1088/1126-6708/2007/11/070}{\doi{10.1088/1126-6708/2007/11/070}},
  \href{http://www.arXiv.org/abs/0709.2092}{\texttt{ arXiv:0709.2092}}.

\bibitem{HERWIG}
\hrefCMSnoop {} {G.~Corcella {et~al.}, ``{HERWIG} 6.1 Release Note'',} (1999).
  \href{http://www.arXiv.org/abs/9912396}{\texttt{ arXiv:9912396}}.

\bibitem{Pythia}
\hrefCMSnoop {} {T.~Sj{\"o}strand, S.~Mrenna, and P.~Skands, ``{PYTHIA} 6.4
  physics and manual'',} \textit{ JHEP} \textbf{ 05} (2006) 026,
  \href{http://dx.doi.org/10.1088/1126-6708/2006/05/026}{\doi{10.1088/1126-6708/2006/05/026}}.

\bibitem{MadGraph}
\hrefCMSnoop {} {J.~Alwall {et~al.}, ``{MadGraph/MadEvent v4}: the new web
  generation'',} \textit{ JHEP} \textbf{ 09} (2007) 028,
  \href{http://dx.doi.org/10.1088/1126-6708/2007/09/028}{\doi{10.1088/1126-6708/2007/09/028}}.

\bibitem{CTEQ6L1}
\hrefCMSnoop {} {J.~Pumplin {et~al.}, ``New Generation of Parton Distributions
  with Uncertainties from Global {QCD} Analysis'',} \textit{ JHEP} \textbf{ 02}
  (2002), no.~07, 012,
  \href{http://dx.doi.org/10.1088/1126-6708/2002/07/012}{\doi{10.1088/1126-6708/2002/07/012}},
  \href{http://www.arXiv.org/abs/arXiv:hep-ph/0201195v3}{\texttt{
  arXiv:arXiv:hep-ph/0201195v3}}.

\bibitem{1107.0330}
\hrefCMSnoop {} {{ {CMS}} Collaboration, ``Measurement of the underlying event
  activity at the {LHC} with $\sqrt{s}= 7$ TeV and comparison with $\sqrt{s} =
  0.9$ {TeV}'',} \textit{ JHEP} (2011) 109,
  \href{http://dx.doi.org/10.1007/JHEP09(2011)109}{\doi{10.1007/JHEP09(2011)109}}.

\bibitem{PFT-09-001}
\href {http://cdsweb.cern.ch/record/1194487} {{ CMS} Collaboration,
  ``Particle--Flow Event Reconstruction in {CMS} and Performance for Jets,
  Taus, and {\MET}'',} CMS Physics Analysis Summary CMS-PAS-PFT-09-001, 2009.

\bibitem{Cacciari:subtraction}
\hrefCMSnoop {} {M.~Cacciari and G.~P. Salam, ``{Pileup subtraction using jet
  areas}'',} \textit{ Phys. Lett. B} \textbf{ 659} (2008) 119,
  \href{http://dx.doi.org/10.1016/j.physletb.2007.09.077}{\doi{10.1016/j.physletb.2007.09.077}},
  \href{http://www.arXiv.org/abs/0707.1378}{\texttt{ arXiv:0707.1378}}.

\bibitem{antikt}
\hrefCMSnoop {} {M.~Cacciari, G.~P. Salam, and G.~Soyez, ``{The anti-$k_t$ jet
  clustering algorithm}'',} \textit{ JHEP} \textbf{ 04} (2008) 063,
  \href{http://dx.doi.org/10.1088/1126-6708/2008/04/063}{\doi{10.1088/1126-6708/2008/04/063}},
  \href{http://www.arXiv.org/abs/0802.1189}{\texttt{ arXiv:0802.1189}}.

\bibitem{Cacciari:fastjet1}
\hrefCMSnoop {} {M.~Cacciari, G.~P. Salam, and G.~Soyez, ``FastJet user
  manual'',} (2011). \href{http://www.arXiv.org/abs/1111.6097}{\texttt{
  arXiv:1111.6097}}.

\bibitem{Cacciari:fastjet2}
\hrefCMSnoop {} {M.~Cacciari and G.~P. Salam, ``{Dispelling the $N^{3}$ myth
  for the $k_t$ jet-finder}'',} \textit{ Phys. Lett. B} \textbf{ 641} (2006)
  57,
  \href{http://dx.doi.org/10.1016/j.physletb.2006.08.037}{\doi{10.1016/j.physletb.2006.08.037}},
  \href{http://www.arXiv.org/abs/hep-ph/0512210}{\texttt{
  arXiv:hep-ph/0512210}}.

\bibitem{cmsJEC}
\hrefCMSnoop {} {{ CMS} Collaboration, ``Determination of Jet Energy
  Calibration and Transverse Momentum Resolution in {CMS}'',} \textit{ JINST}
  \textbf{ 06} (2011) 11002,
  \href{http://dx.doi.org/10.1088/1748-0221/6/11/P11002}{\doi{10.1088/1748-0221/6/11/P11002}}.

\bibitem{CMS-PAS-EGM-10-004}
\href {http://cdsweb.cern.ch/record/1299116} {{ CMS} Collaboration, ``Electron
  Reconstruction and Identification at $\sqrt{s} = 7$ {TeV}'',} CMS Physics
  Analysis Summary CMS-PAS-EGM-10-004, 2010.

\bibitem{CMS-PAS-MUO-10-002}
\href {http://cdsweb.cern.ch/record/1279140} {{ CMS} Collaboration,
  ``Performance of muon identification in pp collisions at $\sqrt{s}$ = 7
  {TeV}'',} CMS Physics Analysis Summary CMS-PAS-MUO-10-002, 2010.

\bibitem{CMS-PAS-BTV-09-001}
\href {http://cdsweb.cern.ch/record/1194494} {{ CMS} Collaboration,
  ``Algorithms for b Jet Identification in {CMS}'',} CMS Physics Analysis
  Summary CMS-PAS-BTV-09-001, 2009.

\bibitem{CMS-PAS-BTV-11-003}
\href {http://cdsweb.cern.ch/record/1421611} {{ CMS} Collaboration,
  ``Measurement of the b-tagging efficiency using \ttbar\ events'',} CMS
  Physics Analysis Summary CMS-PAS-BTV-11-003, 2011.

\bibitem{PhysRevLett.100.242001}
\hrefCMSnoop {} {J.~M. Butterworth {et~al.}, ``Jet Substructure as a New
  {H}iggs-Search Channel at the {Large Hadron Collider}'',} \textit{ {Phys.
  Rev. Lett.}} \textbf{ 100} (2008) 242001,
  \href{http://dx.doi.org/10.1103/PhysRevLett.100.242001}{\doi{10.1103/PhysRevLett.100.242001}}.

\bibitem{tmva}
\hrefCMSnoop {} {A.~Hoecker {et~al.}, ``{TMVA} - toolkit for multivariate data
  analysis with {ROOT}'',} (2007).
\href{http://www.arXiv.org/abs/physics/0703039}{\texttt{
  arXiv:physics/0703039}}.

\bibitem{lumiPAS}
\href {http://cdsweb.cern.ch/record/1376102} {{ CMS} Collaboration, ``Absolute
  Calibration of the {CMS} Luminosity Measurement: {S}ummer 2011 Update'',} CMS
  Physics Analysis Summary CMS-PAS-EWK-11-001, 2011.

\bibitem{Botje:2011sn}
\hrefCMSnoop {} {M.~Botje {et~al.}, ``{The PDF4LHC Working Group Interim
  Recommendations}'',} (2011).
  \href{http://www.arXiv.org/abs/1101.0538}{\texttt{ arXiv:1101.0538}}.

\bibitem{Alekhin:2011sk}
\hrefCMSnoop {} {S.~Alekhin {et~al.}, ``{The PDF4LHC Working Group Interim
  Report}'',} (2011). \href{http://www.arXiv.org/abs/1101.0536}{\texttt{
  arXiv:1101.0536}}.

\bibitem{Lai:2010vv}
\hrefCMSnoop {} {H.~Lai {et~al.}, ``{New parton distributions for collider
  physics}'',} \textit{ Phys. Rev. D} \textbf{ 82} (2010) 074024,
  \href{http://dx.doi.org/10.1103/PhysRevD.82.074024}{\doi{10.1103/PhysRevD.82.074024}},
  \href{http://www.arXiv.org/abs/1007.2241}{\texttt{ arXiv:1007.2241}}.

\bibitem{Martin:2009iq}
\hrefCMSnoop {} {A.~Martin {et~al.}, ``{Parton distributions for the LHC}'',}
  \textit{ Eur. Phys. J. C} \textbf{ 63} (2009) 189,
  \href{http://dx.doi.org/10.1140/epjc/s10052-009-1072-5}{\doi{10.1140/epjc/s10052-009-1072-5}},
  \href{http://www.arXiv.org/abs/0901.0002}{\texttt{ arXiv:0901.0002}}.

\bibitem{Ball:2011mu}
\hrefCMSnoop {} {D.~Richard {et~al.}, ``Impact of Heavy Quark Masses on Parton
  Distributions and LHC Phenomenology''.} Nucl.Phys.B849:296-363,2011, 2011.
\newblock
  \href{http://dx.doi.org/10.1016/j.nuclphysb.2011.03.021}{\doi{10.1016/j.nuclphysb.2011.03.021}}.

\bibitem{HAWK1}
\hrefCMSnoop {} {M.~Ciccolini {et~al.}, ``{Strong and electroweak corrections
  to the production of Higgs+2jets via weak interactions at the LHC}'',}
  \textit{ Phys. Rev. Lett.} \textbf{ 99} (2007) 161803,
  \href{http://dx.doi.org/10.1103/PhysRevLett.99.161803}{\doi{10.1103/PhysRevLett.99.161803}},
  \href{http://www.arXiv.org/abs/0707.0381}{\texttt{ arXiv:0707.0381}}.

\bibitem{HAWK2}
\hrefCMSnoop {} {M.~Ciccolini, A.~Denner, and S.~Dittmaier, ``{Electroweak and
  QCD corrections to Higgs production via vector-boson fusion at the LHC}'',}
  \textit{ Phys. Rev. D} \textbf{ 77} (2008) 013002,
  \href{http://dx.doi.org/10.1103/PhysRevD.77.013002}{\doi{10.1103/PhysRevD.77.013002}},
  \href{http://www.arXiv.org/abs/0710.4749}{\texttt{ arXiv:0710.4749}}.

\bibitem{HAWK3}
\hrefCMSnoop {} {A.~Denner, S.~Dittmaier, S.~Kallweit, and A.~M{\"u}ck,
  ``Electroweak corrections to {H}iggs-strahlung off {W/Z} bosons at the
  {T}evatron and the {LHC} with {HAWK}'',} (2011).
  \href{http://www.arXiv.org/abs/1112.5142}{\texttt{ arXiv:1112.5142}}.

\bibitem{PhysRevD.68.073003}
\hrefCMSnoop {} {M.~L. Ciccolini, S.~Dittmaier, and M.~Kr\"amer, ``Electroweak
  radiative corrections to associated \textit{WH} and \textit{ZH} production at
  hadron colliders'',} \textit{ Phys. Rev. D} \textbf{ 68} (2003) 073003,
  \href{http://dx.doi.org/10.1103/PhysRevD.68.073003}{\doi{10.1103/PhysRevD.68.073003}}.

\bibitem{Grazzini}
\hrefCMSnoop {} {G.~Ferrera, M.~Grazzini, and F.~Tramontano, ``{Associated WH
  production at hadron colliders: a fully exclusive QCD calculation at
  NNLO}'',} (2011). \href{http://www.arXiv.org/abs/1107.1164}{\texttt{
  arXiv:1107.1164}}.

\bibitem{Brein}
\hrefCMSnoop {} {O.~Brein, A.~Djouadi, and R.~Harlander, ``{NNLO QCD}
  corrections to the {H}iggs-strahlung processes at hadron colliders'',}
  \textit{ Phys. Lett. B} \textbf{ 579} (2004) 149,
  \href{http://dx.doi.org/10.1016/j.physletb.2003.10.112}{\doi{10.1016/j.physletb.2003.10.112}},
  \href{http://www.arXiv.org/abs/hep-ph/0307206}{\texttt{
  arXiv:hep-ph/0307206}}.

\bibitem{Read1}
\hrefCMSnoop {} {A.~L. Read, ``Presentation of search results: the {CLs}
  technique'',} \textit{ J. Phys. G: Nucl. Part. Phys.} \textbf{ 28} (2002)
  2693,
  \href{http://dx.doi.org/10.1088/0954-3899/28/10/313}{\doi{10.1088/0954-3899/28/10/313}}.

\bibitem{junkcls}
\hrefCMSnoop {} {T.~Junk, ``{Confidence level computation for combining
  searches with small statistics}'',} \textit{ Nucl. Instrum. Meth. A} \textbf{
  434} (1999) 435,
  \href{http://dx.doi.org/10.1016/S0168-9002(99)00498-2}{\doi{10.1016/S0168-9002(99)00498-2}}.

\bibitem{LHC-HCG}
\href {http://cdsweb.cern.ch/record/1379837} {{ATLAS and CMS Collaborations,
  LHC Higgs Combination Group}, ``Procedure for the {LHC} Higgs boson search
  combination in {S}ummer 2011'',} ATL-PHYS-PUB/CMS NOTE 2011-11, 2011/005,
  2011.

\end{thebibliography}\endgroup

\cleardoublepage \appendix\section{The CMS Collaboration \label{app:collab}}\begin{sloppypar}\hyphenpenalty=5000\widowpenalty=500\clubpenalty=5000\textbf{Yerevan Physics Institute,  Yerevan,  Armenia}\\*[0pt]
S.~Chatrchyan, V.~Khachatryan, A.M.~Sirunyan, A.~Tumasyan
\vskip\cmsinstskip
\textbf{Institut f\"{u}r Hochenergiephysik der OeAW,  Wien,  Austria}\\*[0pt]
W.~Adam, T.~Bergauer, M.~Dragicevic, J.~Er\"{o}, C.~Fabjan, M.~Friedl, R.~Fr\"{u}hwirth, V.M.~Ghete, J.~Hammer\cmsAuthorMark{1}, M.~Hoch, N.~H\"{o}rmann, J.~Hrubec, M.~Jeitler, W.~Kiesenhofer, M.~Krammer, D.~Liko, I.~Mikulec, M.~Pernicka$^{\textrm{\dag}}$, B.~Rahbaran, C.~Rohringer, H.~Rohringer, R.~Sch\"{o}fbeck, J.~Strauss, A.~Taurok, F.~Teischinger, P.~Wagner, W.~Waltenberger, G.~Walzel, E.~Widl, C.-E.~Wulz
% ...
\vskip\cmsinstskip
\dag:~Deceased\\
1:~~Also at CERN, European Organization for Nuclear Research, Geneva, Switzerland\\
% ...

\end{sloppypar}
\end{document}